\documentclass[pre,eqsecnum,tighten]{revtex4}

\usepackage{amssymb,amsmath}

\usepackage{epsfig}

\usepackage{graphicx}

\usepackage{bm}

\begin{document}
\title{Enskog Theory for Polydisperse Granular Mixtures II. Sonine Polynomial Approximation}
\author{Vicente Garz\'{o}\footnote[1]{Electronic address: vicenteg@unex.es}}
\affiliation{Departamento de F\'{\i}sica, Universidad de
Extremadura, E-06071 Badajoz, Spain}
\author{Christine M. Hrenya
\footnote[3]{Electronic address: hrenya@colorado.edu}}
\affiliation{Department of Chemical and Biological Engineering,
University of Colorado, Boulder, CO 80309}
\author{James W. Dufty\footnote[2]
{Electronic address: dufty@phys.ufl.edu}} \affiliation{Department of
Physics, University of Florida, Gainesville, Florida 32611}

\begin{abstract}
The linear integral equations defining the Navier-Stokes (NS) transport coefficients for polydisperse granular
mixtures of smooth inelastic hard disks or spheres are solved by using the leading terms in a Sonine polynomial
expansion. Explicit expressions for all the NS transport coefficients are given in terms of the sizes, masses,
compositions, density and restitution coefficients. In addition, the cooling rate is also evaluated to first
order in the gradients. The results hold for arbitrary degree of inelasticity and are not limited to specific
values of the parameters of the mixture. Finally, a detailed comparison between the derivation of the current
theory and previous theories for mixtures is made, with attention paid to the implication of the various
treatments employed to date.
\end{abstract}

 \draft
\date{\today}
\maketitle

\section{Introduction}

\label{sec1}

In the first portion \cite{DGH06} of this two-paper effort on the development of a kinetic-theory-based
description of mixtures, a rigorous Chapman-Enskog (CE) expansion of the revised Enskog theory (RET) for an $s$
-component mixture of inelastic hard spheres was carried out to first (Navier-Stokes) order in spatial
gradients. The expansion was performed about a homogeneous cooling state (HCS), which is the zeroth-order
solution of the kinetic equation for the single-particle velocity distribution function $f_{i}$ of species $i$.
Unlike previous theories for mixtures which derive from expansions about an elastic base state \cite
{Jenkins87,Jenkins89,Zamankhan95,Arnarson98,Willits99, Huilin01,Rahaman03,Iddir05,Serero06}, this CE expansion
does not impose any constraints on the level of dissipation. The resulting theory is thus expected to be
applicable to a wide range of restitution coefficients. Furthermore, because the derivation used the RET as its
starting point (as opposed to the Boltzmann equation as used in Ref.\ \cite{Garzo02}), the results are expected
to be applicable to dilute and moderately dense systems. The theory is not expected to be applicable to systems
dense enough for ring collisions to play a significant role, since such velocity correlations are not accounted
for in the RET.

This formally exact analysis for Navier-Stokes order hydrodynamics, reported in the companion paper, resulted in
integral-differential equations for the zeroth-order $f_{i}^{(0)}$ and first-order $f_{i}^{(1)}$ distribution
functions as well as integral expressions (in terms of $f_{i}^{(0)}$ and $f_{i}^{(1)}$) for each of the
equations of state (cooling rate and pressure) and the transport coefficients
$\{D_{ij},D_{i}^{T},D_{ij}^{F},\eta ,\kappa ,\lambda ,D_{q,ij},L_{ij}\}$. Of these, only the pressure could be
directly evaluated and cast in algebraic expressions of the macroscopic (hydrodynamic) variables. Hence, in this
second part of the work, approximate methods are used to obtain algebraic equations for the kinetic and
collisional contributions to the cooling rate and transport coefficients. In particular, the equations for
$f_{i}^{(0)}$ have previously been solved \cite{Garzo99} via a combination of scaling arguments and an
approximation to the distribution function based on leading-order Sonine polynomials. Recent results derived for
binary granular mixtures at low-density \cite{Garzo02,GMD06} have shown that the influence of the non-Gaussian
(higher-order) corrections of $f_{i}^{(0)}$ to the transport coefficients is in general negligible, except for
quite large values of dissipation and velocities. For this reason, we use leading order in the Sonine polynomial
expansion (Maxwellians) at different temperatures to evaluate integrals over the distributions $f_{i}^{(0)}$.
The solutions for the $f_{i}^{(1)}$ are also found using a truncated polynomial expansion. These forms for $
f_{i}^{(0)}$ and $f_{i}^{(1)}$ are then used to obtain practical expressions for the cooling rate and transport
coefficients. Namely, the resulting, algebraic constitutive equations depend explicitly on the hydrodynamic
variables only and not on the distribution functions of the mixture. Collision integrals are reduced to Gaussian
forms and evaluated using standard integration techniques; a computer package of symbolic calculations
(Mathematica) was used to check their analytical evaluation. Wherever possible, limiting values of the equations
of state and transport coefficients were verified by comparison with previously published works in the cases of
mechanically equivalent particles \cite{Garzo99a,Lutsko05}, binary mixtures at low-density \cite{Garzo02,GM06},
and simple shear flow states for binary mixtures at moderate density \cite{GM03}. All analytic calculations of
both papers were performed independently and cross-checked.

Finally, as mentioned in the Introduction of the companion paper, a myriad of treatments has appeared in the
literature for the derivation of kinetic-theory-based descriptions of granular mixtures. To help put the current
effort in the context of previous contributions, Sec.\ VI contains a breakdown of the theoretical contributions
to date, along with a critical discussion of the ramifications of each treatment.

\section{Euler Order Parameters}

The hydrodynamic equations to first order in the spatial gradients (Euler order) have as unknown
(phenomenological) parameters the pressure and the cooling rate,
\begin{equation}
p=p(T,\{n_{i}\}),\hspace{0.25in}\zeta (T,\{n_{i}\})=\zeta ^{(0)}(T,\{n_{i}\})+\zeta _{U}(T,\{n_{i}\})\nabla
\cdot \mathbf{U}. \label{1.1}
\end{equation}
The pressure $p(T,\{n_{i}\})$ has been evaluated exactly in the previous paper \cite{DGH06} [Eqs.\ (7.18) and
(7.19)] with the result
\begin{equation}
p(T,\{n_{i}\})=nT+\frac{\pi ^{d/2}}{d\Gamma \left( \frac{d}{2}\right) } \sum_{i=1}^{s}\sum_{j=1}^{s}\mu
_{ji}\sigma _{ij}^{d}\chi _{ij}^{(0)}n_{i}n_{j}(1+\alpha _{ij})T_{i},  \label{1.2}
\end{equation}
where henceforth, for the sake of convenience, we will use the notation $T_i\equiv T_i^{(0)}$ with $T_i^{(0)}$
defined by Eq.\ (7.5) of Ref.\ \cite{DGH06}. Detailed forms for the functions $\chi _{ij}^{(0)}$ are discussed
in Appendix \ref{appC}. The leading order cooling rate, $\zeta ^{(0)}(T,\{n_{i}\}),$ also is given there by Eq.
(7.2) as an integral over $ f_{i}^{(0)}$. For reasons just mentioned above, these integrals are performed using
the leading order Sonine approximation for $f_{i}^{(0)}$
\begin{equation}
f_{i}^{(0)}(\mathbf{V})\rightarrow f_{i,M}(\mathbf{V})=n_{i}\left( \frac{ m_{i}}{2\pi T_{i}}\right) ^{d/2}\exp
\left( -\frac{m_{i}V^{2}}{2T_{i}} \right) ,  \label{1.9a}
\end{equation}
and $\mathbf{V}=\mathbf{v}-\mathbf{U}$ is the peculiar velocity. The result is \cite{Garzo99a,GM06}
\begin{eqnarray}
\zeta ^{(0)}(T,\{n_{i}\}) &=&\zeta _{i}^{(0)}(T,\{n_{i}\})  \notag \\
&=&\frac{4\pi ^{(d-1)/2}}{d\Gamma \left( \frac{d}{2}\right) } v_{0}\sum_{j=1}^{s}\chi _{ij}^{(0)}n_{j}\mu
_{ji}\sigma _{ij}^{d-1}\left( \frac{\theta _{i}+\theta _{j}}{\theta _{i}\theta _{j}}\right) ^{1/2}(1+\alpha
_{ij})\left[ 1-\frac{\mu _{ji}}{2}(1+\alpha _{ij})\frac{ \theta _{i}+\theta _{j}}{\theta _{j}}\right] ,
\nonumber\\ \label{1.3}
\end{eqnarray}
where $n=\sum_{i}n_{i}$ is the total density, $v_{0}(t)=\sqrt{2T/m}$ is a thermal velocity,
$m=(\sum_{j}m_{j})/s$, $\mu _{ij}=m_{i}/(m_{i}+m_{j})$, and $\theta _{i}=m_{i}T/mT_{i}$. The $\zeta
_{i}^{(0)}(T,\{n_{i}\})$ are species cooling rates, measuring the decrease of kinetic energy of each species.
This is an implicit definition of $\zeta ^{(0)}(T,\{n_{i}\})$. The equality of cooling rates for all $i$ gives
$s-1$ equations for the species temperatures $T_{i}$ in terms of $T$. A final equation is given by the condition
that the total kinetic energy is the sum of species energies $nT=\sum_{i=1}^{s}n_{i}T_{i}$. With the species
temperatures determined as functions of $T$ and $\{n_{i}\}$, Eq. (\ref{1.3}) gives the cooling rate $\zeta
^{(0)}(T,\{n_{i}\})$. In all of the following expressions, it is understood that\ the $T_{i}(T,\{n_{i}\})$ have
been determined in this way.

At first order in gradients, there is a contribution to the cooling rate
from $\nabla \cdot \mathbf{U}$. The proportionality coefficient $\zeta _{U}$
is a new transport coefficient for granular fluids. Two different
contributions can be identified
\begin{equation}
\zeta _{U}=\zeta ^{(1,0)}+\zeta ^{(1,1)}.  \label{1.4}
\end{equation}
The coefficient $\zeta ^{(1,0)}$ is given by an integral over $f_{i}^{(0)}$ and its velocity derivative. It has
been evaluated explicitly in Appendix F of Ref.\ \cite{DGH06} with the result
\begin{equation}
\zeta ^{(1,0)}=-\frac{3}{nT}\frac{\pi ^{d/2}}{d^{2}\Gamma \left( \frac{d}{2} \right)
}\sum_{i=1}^{s}\sum_{j=1}^{s}n_{i}n_{j}\mu _{ji}\sigma _{ij}^{d}\chi _{ij}^{(0)}(1-\alpha _{ij}^{2})T_{i} .
\label{1.5}
\end{equation}

The second term of Eq.\ (\ref{1.4}), $\zeta ^{(1,1)},$ is given by
\begin{equation}
\zeta ^{(1,1)}=\frac{1}{nT}\frac{\pi ^{(d-1)/2}}{d\Gamma \left( \frac{d+3}{2} \right)
}\sum_{i=1}^{s}\sum_{j=1}^{s}\sigma _{ij}^{d-1}\chi _{ij}^{(0)}m_{j}\mu _{ij}(1-\alpha _{ij}^{2})\int
d\mathbf{v}_{1}\,\int d \mathbf{v}_{2}\,g^{3}f_{i}^{(0)}(\mathbf{V}_{1})\mathcal{D}_{j}(\mathbf{V} _{2}),
\label{1.6}
\end{equation}
where the unknown functions $\mathcal{D}_{i}(\mathbf{V})$ are the solutions
to the linear integral equations
\begin{equation}
\left( \left( \mathcal{L}+\frac{1}{2}\zeta ^{(0)}\right) \mathcal{D}\right) _{i}+\frac{1}{2}\zeta
^{(1,1)}(\mathcal{D}_{i})\nabla _{\mathbf{V}}\cdot \left( \mathbf{V}f_{i}^{(0)}\right) =\overline{D}_{i}.
\label{1.7}
\end{equation}
Here, $\nabla_{\bf V}\equiv \partial/\partial {\bf V}$ and the linear operator $\mathcal{L}$ is
\begin{equation}
\left( \mathcal{L}X\right) _{i}=\frac{1}{2}\zeta ^{(0)}\nabla _{\mathbf{V} }\cdot \left( \mathbf{V}X_{i}\right)
+\left( LX\right) _{i},  \label{1.17a}
\end{equation}
where $L$ is the linearized Enskog collision operator
\begin{equation}
\left( LX\right) _{i}=-\sum_{j=1}^{s}\left( J_{ij}^{(0)}\left[ \mathbf{v} _{1}\mid X_{i},f_{j}^{(0)}\right]
+J_{ij}^{(0)}\left[ \mathbf{v}_{1}\mid f_{i}^{(0)},X_{j}\right] \right) ,  \label{1.17b}
\end{equation}
\begin{eqnarray}
J_{ij}^{(0)}\left[ \mathbf{v}_{1}\mid X_{i},f_{j}\right] &\equiv &\chi _{ij}^{(0)}\left( \sigma _{ij};\left\{
n_{i}\left( \mathbf{r}_{1},t\right) \right\} \right) \sigma _{ij}^{d-1}\int d\mathbf{v}_{2}\int d\widehat{
\boldsymbol{\sigma }}\Theta (\widehat{\boldsymbol{\sigma }}\cdot \mathbf{g}
_{12})(\widehat{\boldsymbol{\sigma }}\cdot \mathbf{g}_{12})  \notag \\
&&\times \left[ \alpha _{ij}^{-2}X_{i}(\mathbf{V}_{1}^{\prime \prime })f_{j}( \mathbf{V}_{2}^{\prime \prime
})-X_{i}(\mathbf{V}_{1})f_{j}(\mathbf{V}_{2}) \right] ,  \label{1.17c}
\end{eqnarray}
and $\mathbf{v}_{1}^{\prime \prime },\mathbf{v}_{2}^{\prime \prime }$ are the precollision velocities associated
with $\mathbf{v}_{1},\mathbf{v}_{2}$ for the colliding pair of species $i$ and $j$. The same linear operator $
\mathcal{L}$ appears below in the definitions of all other transport coefficients as well. The inhomogeneity of
Eq.\ (\ref{1.7}), $\overline{D} _{i}(T,\{n_{i}\})$, is given by
\begin{eqnarray}
\overline{D}_{i} &=&\left( \frac{1}{d}\left( 1-\frac{p}{nT}\right) -\frac{1}{ 2}\zeta ^{(1,0)}\right) \nabla
_{\mathbf{V}}\cdot \left( \mathbf{V} f_{i}^{(0)}\right) -f_{i}^{(0)}+\sum_{j=1}^{s}n_{j}\frac{\partial
f_{i}^{(0)}
}{\partial n_{j}}  \notag \\
&&+\frac{1}{d}\sum_{j=1}^{s}\mathcal{K}_{ij,\beta }[\partial _{V_{\beta
}}f_{j}^{(0)}],  \label{1.8}
\end{eqnarray}
and the operator $\mathcal{K}_{ij,\beta }$ is given by (\ref{b0}) of Appendix \ref{appB}. An approximate
solution to this integral equation is obtained by using $f_{i}^{(0)}$, $p$, $\zeta ^{(0)}$, and $\zeta ^{(1,0)}$
as determined above, and the leading order term in an expansion of $\mathcal{ D}_{i}(\mathbf{V})$ in a complete
set of Sonine polynomials
\begin{equation}
\mathcal{D}_{i}(\mathbf{V})\rightarrow e_{i,D}f_{i,M}(\mathbf{V})F_{i}( \mathbf{V}).\hspace{0.25in}  \label{1.9}
\end{equation}
The lowest order Sonine polynomial contributing in this case is
\begin{equation}
F_{i}(\mathbf{V})=\left( \frac{m_{i}}{2T_{i}}\right) ^{2}V^{4}-\frac{d+2}{2}
\frac{m_{i}}{T_{i}}V^{2}+\frac{d(d+2)}{4},  \label{1.10}
\end{equation}
as determined by the fact that $\mathcal{D}_{i}(\mathbf{V})$ is a scalar and orthogonal to $1$ and $V^{2}$.
Finally, $e_{i,D}$ is the projection of $ \mathcal{D}_{i}$ along $F_{i}$
\begin{equation}
e_{i,D}=\frac{2}{d(d+2)}\frac{1}{n_{i}}\int \;d\mathbf{v}\;\mathcal{D}_{i}( \mathbf{V})F_{i}(\mathbf{V}).
\label{1.11}
\end{equation}
Note that the polynomials are defined for each species in terms of the weight factor $f_{i,M}(\mathbf{V})$
characterized by the mass and temperature of that species \cite{Garzo02}. The coefficients $ e_{i,D}$ are
determined by substituting (\ref{1.9}) into the integral equation (\ref{1.7}), multiplying by
$F_{i}(\mathbf{V})$, and integrating over $\mathbf{V}$ (this assures that the integral equation is exactly
satisfied in the subspace spanned $F_{i}$) to get the algebraic equations
\begin{equation}
\sum_{j=1}^{s}\left( \psi _{ij}-\frac{3}{2}\zeta ^{(0)}\delta _{ij}\right)
e_{j,D}=\overline{e}_{i,D}.  \label{1.13}
\end{equation}
The collision frequencies $\psi _{ij}$ are defined by
\begin{equation}
\psi _{ii}=-\frac{2}{d(d+2)}\frac{1}{n_{i}}\left( \sum_{j=1}^{s}\int d
\mathbf{v}\,F_{i}J_{ij}^{(0)}[f_{i,M}F_{i},f_{j}^{(0)}]+\int d\mathbf{v}
\,F_{i}J_{ii}^{(0)}[f_{i}^{(0)},f_{i,M}F_{i}]\right) ,  \label{1.14}
\end{equation}
\begin{equation}
\psi _{ij}=-\frac{2}{d(d+2)}\frac{1}{n_{i}}\int d\mathbf{v} \,F_{i}J_{ij}^{(0)}[f_{i}^{(0)},f_{j,M}F_{j}],\quad
(i\neq j).  \label{1.15}
\end{equation}
The inhomogeneity on the right-hand side of (\ref{1.13}) is computed in
Appendix \ref{appB} with the result
\begin{eqnarray}
\overline{e}_{i,D} &=&\frac{1}{d}\sum_{j=1}^{s}\int d\mathbf{v}\,F_{i}( \mathbf{V})\mathcal{K}_{ij,\gamma
}[\partial _{V_{\gamma }}f_{j}^{(0)}]
\notag \\
&=&\frac{\pi ^{d/2}}{4d\Gamma \left( \frac{d}{2}\right) } \sum_{j=1}^{s}n_{i}n_{j}\chi _{ij}^{(0)}\sigma
_{ij}^{d}\mu _{ji}(1+\alpha
_{ij})\left[ 8(d+2)(\mu _{ij}-1)\right.  \notag \\
&&+4(13+2d+9\alpha _{ij})\mu _{ji}-48\mu _{ji}^{2}\theta _{j}^{-1}(\theta
_{i}+\theta _{j})(1+\alpha _{ij})^{2}  \notag \\
&&\left. +15\mu _{ji}^{3}\theta _{j}^{-2}(\theta _{i}+\theta
_{j})^{2}(1+\alpha _{ij})^{3}\right] .  \label{1.16}
\end{eqnarray}
The explicit form of the collision frequencies $\psi _{ij}$ are displayed in
Appendix \ref{appA}, so the coefficients $e_{i,D}$ can be determined by
solving the algebraic Eqs.\ (\ref{1.13}) with all coefficients determined.

Finally, $\zeta ^{(1,1)}$ is given in terms of $e_{i,D}$ by substitution of
Eq.\ (\ref{1.9}) into (\ref{1.6}), with the result
\begin{equation}
\zeta ^{(1,1)}=e_{i,D}\frac{3\pi ^{(d-1)/2}}{4d\Gamma \left( \frac{d}{2} \right)
}\frac{v_{0}^{3}}{nT}\sum_{i=1}^{s}\sum_{j=1}^{s}\sigma _{ij}^{d-1}\chi _{ij}^{(0)}m_{j}\mu _{ij}(1-\alpha
_{ij}^{2})\theta _{i}^{-3/2}\theta _{j}^{1/2}(\theta _{i}+\theta _{j})^{-1/2}.  \label{1.17}
\end{equation}

\section{Navier-Stokes Order Transport Coefficients}
\label{sec2}

The hydrodynamic equations to second order in the spatial gradients have contributions to the mass flux,
pressure tensor, and heat flux from the Chapman-Enskog expansion proportional to gradients of the species
densities, temperature, and flow velocity and to an external applied force (of the same order of magnitude as
the gradients). For fluid symmetry, this leads to $(s-1)(2s+1)$ transport coefficients for the mass flux, the
shear and bulk viscosity for the pressure tensor, and $2s^{2}+1$ coefficients for the heat flux. The mass flux
coefficients are determined from integral equations in the same way as $\zeta ^{(1,1)}$, just described. The
pressure tensor and heat fluxes have ``kinetic'' and ``collisional'' transfer parts, i.e. of the forms
\begin{equation*}
\omega =\omega ^{k}+\omega ^{c}.
\end{equation*}
The contributions $\omega ^{k}$ are determined from integral equations like that for $\zeta ^{(1,1)}$; the
collisional transfer contributions have been reduced in the previous paper to specific integrals over
$f_{i}^{(0)}$ and can be computed explicitly by using the multi-temperature Maxellians (\ref{1.9a}). The
analysis is similar to that for the cooling rate and many of the details are transferred to the Appendices.

\subsection{Mass Flux}

To first order in the spatial gradients, the mass flux $\mathbf{j}_{0i}^{(1)}$ is given by
\begin{equation}
\mathbf{j}_{0i}^{(1)}=-\sum_{j=1}^{s}\frac{m_{i}m_{j}n_{j}}{\rho } D_{ij}\nabla \ln n_{j}-\rho D_{i}^{T}\nabla
\ln T-\sum_{j=1}^{s}D_{ij}^{F} \mathbf{F}_{j},  \label{2.1}
\end{equation}
where $\rho =\sum_{i}\rho _{i}=\sum_{i}m_{i}n_{i}$ is the total mass density
of the mixture, $D_{ij}$ are the mutual diffusion coefficients, $D_{i}^{T}$
are the thermal diffusion coefficients and $D_{ij}^{F}$ are the mobility
coefficients. These transport coefficients are defined as
\begin{equation}
D_{ij}=-\frac{\rho }{dm_{j}n_{j}}\int d\mathbf{v}\,\mathbf{V}\cdot
\boldsymbol{\mathcal{B}}_{i}^{j}(\mathbf{V}),  \label{2.2}
\end{equation}
\begin{equation}
D_{i}^{T}=-\frac{m_{i}}{d\rho }\int d\mathbf{v}\,\mathbf{V}\cdot \boldsymbol{ \mathcal{A}}_{i}(\mathbf{V}),
\label{2.3}
\end{equation}
\begin{equation}
D_{ij}^{F}=-\frac{m_{i}}{d}\int d\mathbf{v}\,\mathbf{V}\cdot \boldsymbol{ \mathcal{E}}_{i}^{j}(\mathbf{V}).
\label{2.3.1}
\end{equation}
The functions $\mathcal{A}_{i}(\mathbf{V})$, $\mathcal{B}_{i}^{j}(\mathbf{V} ) $, and
${\boldsymbol{\mathcal{E}}}_{i}^{j}(\mathbf{V})$ verify the linear integral equations
\begin{equation}
\left( \left( \mathcal{L}-\frac{1}{2}\zeta ^{(0)}\right) \boldsymbol{ \mathcal{A}}\right) _{i}=\mathbf{A}_{i},
\label{2.3.2}
\end{equation}
\begin{equation}
\left( \mathcal{L}\boldsymbol{\mathcal{B}} ^{j}\right) _{i}-n_{j}\frac{\partial \zeta ^{(0)}}{
\partial n_{j}}\boldsymbol{\mathcal{A}}_{i}=\mathbf{B}_{i}^{j},
\label{2.3.3}
\end{equation}
\begin{equation}
\left( \left( \mathcal{L}+\zeta ^{(0)}\right) \boldsymbol{\mathcal{E}} ^{j}\right) _{i}=\mathbf{E}_{i}^{j},
\label{2.3.4}
\end{equation}
where the linear operator $\mathcal{L}$ and the inhomogeneous terms $\mathbf{ A}_{i}$, $\mathbf{B}_{i}^{j}$, and
$\mathbf{E}_{i}^{j}$ are given by Eqs. (6.16), (6.17), and (6.20) of the preceding paper \cite{DGH06}. The
approximate solutions to these equations are obtained in the same manner as that for ( \ref{1.7}). The leading
order Sonine polynomial in all of these cases is $ \mathbf{V}$, since all are vectors
\begin{equation}
\boldsymbol{\mathcal{A}}_{i}(\mathbf{V})\rightarrow -\frac{\rho }{n_{i}T_{i}}
D_{i}^{T}f_{i,M}(\mathbf{V})\mathbf{V},  \label{2.4}
\end{equation}
\begin{equation}
\boldsymbol{\mathcal{B}}_{i}^{j}(\mathbf{V})\rightarrow -\frac{ m_{i}m_{j}n_{j}}{\rho
n_{i}T_{i}}D_{ij}f_{i,M}(\mathbf{V})\mathbf{V}, \label{2.5}
\end{equation}
\begin{equation}
\boldsymbol{\mathcal{E}}_{i}^{j}(\mathbf{V})\rightarrow -\frac{1}{n_{i}T_{i}}
D_{ij}^{F}f_{i,M}(\mathbf{V})\mathbf{V},  \label{2.5.1}
\end{equation}
The coefficients are now projections of $\boldsymbol{\mathcal{A}}_{i}$, $\boldsymbol{\mathcal{B}}_{i}^{j}$, and
$\boldsymbol{\mathcal{E}}_{i}^{j}$ along $\mathbf{V}$, which are identified in terms of the transport
coefficients through Eqs.\ (\ref {2.2})--(\ref{2.3.1}). Multiplication of Eqs.\ (\ref{2.3.2})--(\ref{2.3.4}) by
$m_{i}\mathbf{V}$ and integrating over the velocity yields the algebraic equations determining all mass flux
transport coefficients
\begin{equation}
\sum_{j=1}^{s}(\nu _{ij}-\zeta ^{(0)}\delta _{ij})D_{j}^{T}=-\frac{1}{d\rho } \int
d\mathbf{v}\,m_{i}\mathbf{V}\cdot \mathbf{A}_{i}(\mathbf{V}), \label{2.7}
\end{equation}
\begin{equation}
\sum_{\ell =1}^{s}\left( \nu _{i\ell }-\frac{1}{2}\zeta ^{(0)}\delta _{i\ell }\right)\frac{m_\ell}{m_i} D_{\ell
j}-\frac{\rho ^{2}}{m_{i}m_{j}}\frac{\partial \zeta ^{(0)}}{
\partial n_{j}}D_{i}^{T}=-\frac{1}{d}\frac{\rho }{m_{i}m_{j}n_{j}}\int d
\mathbf{v}\,m_{i}\mathbf{V}\cdot \mathbf{B}_{i}^{j}(\mathbf{V}),  \label{2.8}
\end{equation}
\begin{equation}
\sum_{\ell =1}^{s}\left( \nu _{i\ell }+\frac{1}{2}\zeta ^{(0)}\delta _{i\ell }\right)\frac{m_\ell}{m_i} D_{\ell
j}^{F}=-\frac{1}{d}\int d\mathbf{v}\,m_{i}\mathbf{V}\cdot \mathbf{E}_{i}^{j}(\mathbf{V}).  \label{2.8.1}
\end{equation}
The new collision frequencies $\nu _{ij}$ are
\begin{equation}
\nu _{ii}=-\frac{1}{dn_{i}T_{i}}\sum_{j\neq i}^{s}\int d\mathbf{v}\,m_{i} \mathbf{V}\cdot
J_{ij}^{(0)}[f_{i,M}\mathbf{V},f_{j}^{(0)}],  \label{2.9}
\end{equation}
\begin{equation}
\nu _{ij}=-\frac{1}{dn_{j}T_{j}}\int d\mathbf{v}\,m_{i}\mathbf{V}\cdot
J_{ij}^{(0)}[f_{i}^{(0)},f_{j,M}\mathbf{V}],\quad (i\neq j).  \label{2.10}
\end{equation}
Note that the self-collision terms of $\nu _{ii}$ arising from $ J_{ii}^{(0)}[f_{i,M}\mathbf{V},f_{i}^{(0)}]$ do
not occur in (\ref{2.9}) since these conserve momentum for species $i$. The above collision frequencies were
already evaluated in the Boltzmann limit (except for the factors $\chi _{ij}^{(0)})$ \cite{Garzo02,GM06}. The
details will not repeated here and only the results are quoted in Appendix \ref{appA}. The integrals on the
right hand side of Eqs.\ (\ref{2.7}), (\ref{2.8}), and (\ref {2.8.1}) can be performed using the definitions of
$\mathbf{A}_{i}$, $ \mathbf{B}_{i}^{j}$, and $\mathbf{E}_{i}^{j}$. The result is
\begin{equation}
\int d\mathbf{v}\,m_{i}\mathbf{V}\cdot \mathbf{A}_{i}=dp\frac{\rho _{i}}{ \rho }\left( 1-\frac{\rho
n_{i}T_{i}}{\rho _{i}p}\right) +\frac{1}{2} \sum_{j=1}^{s}\int d\mathbf{v}\,m_{i}V_{\gamma
}\mathcal{K}_{ij,\gamma }[\nabla _{\mathbf{V}}\cdot (\mathbf{V}f_{j}^{(0)})],  \label{2.11}
\end{equation}
\begin{eqnarray}
\int d\mathbf{v}\,m_{i}\mathbf{V}\cdot \mathbf{B}_{i}^{j} &=&-dn_{j}\frac{
\partial }{\partial n_{j}}(n_{i}T_{i})+d\frac{\rho _{i}}{\rho }n_{j}\frac{
\partial p}{\partial n_{j}}  \notag  \label{2.12} \\
&&-\sum_{\ell =1}^{s}\int d\mathbf{v}\,m_{i}V_{\gamma }\mathcal{K}_{i\ell
,\gamma }\left[ \left( n_{j}\partial _{n_{j}}+\frac{1}{2}\left(
n_{j}\partial _{n_{j}}\ln \chi _{i\ell }^{(0)}+I_{i\ell j}\right) \right)
f_{\ell }^{(0)}\right] ,  \notag \\
&&
\end{eqnarray}
\begin{equation}
\int d\mathbf{v}\,m_{i}\mathbf{V}\cdot \mathbf{E}_{i}^{j}=d\frac{n_{i}m_{i}}{ m_{j}}\left( \delta
_{ij}-\frac{n_{j}m_{j}}{\rho }\right) .  \label{2.11.1}
\end{equation}
The integrals appearing in Eqs.\ (\ref{2.11}) and (\ref{2.12}) that involve the operator $\mathcal{K}_{ij,\gamma
}[X]$ have been evaluated in Appendix \ref{appB}. They are given by Eqs.\ (\ref{b7}), (\ref{b8}), and
(\ref{b9}). With these results, Eqs.\ (\ref{2.7}), (\ref{2.8}), and (\ref{2.8.1}) become
\begin{equation}
\sum_{j=1}^{s}(\nu _{ij}-\zeta ^{(0)}\delta _{ij})D_{j}^{T}=-\frac{p\rho _{i} }{\rho ^{2}}\left( 1-\frac{\rho
n_{i}T_{i}}{\rho _{i}p}\right) +\frac{\pi ^{d/2}}{d\Gamma \left( \frac{d}{2}\right) }\frac{n_{i}}{\rho }
\sum_{j=1}^{s}n_{j}\mu _{ij}\chi _{ij}^{(0)}\sigma _{ij}^{d}T_{j}(1+\alpha _{ij}),  \label{2.13}
\end{equation}
\begin{eqnarray}
\sum_{\ell =1}^{s}\left( \nu _{i\ell }-\frac{1}{2}\zeta ^{(0)}\delta _{i\ell }\right)\frac{m_\ell}{m_i} D_{\ell
j} &=&\frac{\rho ^{2}}{m_{i}m_{j}}\frac{\partial \zeta ^{(0)}}{\partial n_{j}}D_{i}^{T}+\frac{\rho
}{m_{i}m_{j}}\frac{\partial }{
\partial n_{j}}(n_{i}T_{i})-\frac{n_{i}}{m_{j}}\frac{\partial p}{\partial
n_{j}}  \notag  \label{2.14} \\
&&+\frac{\pi ^{d/2}}{d\Gamma \left( \frac{d}{2}\right) }\frac{\rho n_{i}}{ m_{j}}\sum_{\ell =1}^{s}\chi _{i\ell
}^{(0)}\;\sigma _{i\ell }^{d}\;\mu _{\ell i}(1+\alpha _{i\ell })\left\{ \left( \frac{T_{i}}{m_{i}}+\frac{
T_{\ell }}{m_{\ell }}\right) \right.  \notag \\
&&\times \left. \left[ \delta _{j\ell }+\frac{1}{2}\frac{n_\ell}{n_j}\left( n_{j}\frac{
\partial }{\partial n_{j}}\ln \chi _{i\ell }^{(0)}+I_{i\ell j}\right) \right]
+\frac{n_{\ell }T_{\ell }}{m_{\ell }}\frac{\partial }{\partial n_{j}}\ln
\gamma _{\ell }\right\} ,  \notag \\
&&
\end{eqnarray}
\begin{equation}
\sum_{\ell =1}^{s}\left( \nu _{i\ell }+\frac{1}{2}\zeta ^{(0)}\delta _{i\ell }\right)\frac{m_\ell}{m_i} D_{\ell
j}^{F}=-\frac{n_{i}m_{i}}{m_{j}}\left( \delta _{ij}-\frac{ n_{j}m_{j}}{\rho }\right) ,  \label{2.14.1}
\end{equation}
where $\gamma _{i}\equiv T_{i}/T$ is the temperature ratio. An explicit form for $I_{i\ell j}$ is chosen in
Appendix \ref{appC}. The solution of the set of algebraic equations (\ref{2.13})--(\ref{2.14.1}) gives the
dependence of the coefficients $D_{ij}$, $D_{i}^{T}$, and $D_{ij}^{F}$ on the restitution coefficients $\alpha
_{ij}$ and the composition, the density, and the sizes and masses of the constituents of the mixture.

\subsection{Pressure Tensor}
\label{sec3}

The constitutive equation for the pressure tensor $P_{\alpha \beta }^{(1)}$,
proportional to the velocity gradients, is
\begin{equation}
P_{\alpha \beta }^{(1)}=-\eta \left( \partial r_{\alpha }U_{\beta }+\partial
r_{\beta }U_{\alpha }-\frac{2}{d}\delta _{\alpha \beta }\mathbf{\nabla \cdot
u}\right) -\kappa \delta _{\alpha \beta }\mathbf{\nabla \cdot u}.
\label{3.1}
\end{equation}
Here, $\eta $ is the shear viscosity coefficient and $\kappa $ is the bulk viscosity. The coefficient $\eta $
has kinetic and collisional contributions while $\kappa $ only has a collisional contribution $\kappa ^{c}$ (and
so, vanishes for dilute gases)
\begin{equation}
\eta =\eta ^{k}+\eta ^{c},\hspace{0.25in}\kappa =\kappa ^{c}.  \label{3.1a}
\end{equation}
The collisional transfer contributions have been analyzed in the preceding paper \cite{DGH06} [Eqs. (7.21) and
(7.22)]. These expressions reduce to those previously derived in the monodisperse case \cite
{Garzo99a,Lutsko05}, and in the case of binary mixtures of hard spheres \cite {GM03}. The integrals over
$f_{i}^{(0)}(\mathbf{V})$ are easily performed in the multi-temperature Maxwellian approximation (\ref{1.9a})
with the results
\begin{equation}
\kappa ^{c}=\frac{\pi ^{(d-1)/2}}{d^2\Gamma \left( \frac{d}{2}\right)}
\sum_{i=1}^{s}\sum_{j=1}^{s}\frac{m_{i}m_{j}}{m_{i}+m_{j}} n_{i}n_{j}v_{0}\sigma _{ij}^{d+1}\chi
_{ij}^{(0)}(1+\alpha _{ij})\left( \frac{\theta _{i}+\theta _{j}}{\theta _{i}\theta _{j}}\right) ^{1/2},
\label{3.8}
\end{equation}
\begin{equation}
\eta ^{c}=\frac{2\pi ^{d/2}}{\Gamma \left( \frac{d}{2}\right) }
\frac{1}{d(d+2)}\sum_{i=1}^{s}\sum_{j=1}^{s}n_{j}\sigma _{ij}^{d}\chi _{ij}^{(0)}\mu _{ji}(1+\alpha _{ij})\eta
_{i}^{k}+\frac{d}{d+2}\kappa ^{c}.  \label{3.9}
\end{equation}

\subsubsection{Kinetic contribution $\protect\eta ^{k}$}
\label{subsec3.1}

As noted above, there is no kinetic part of the bulk viscosity, $\kappa
^{k}=0$, so $\kappa $ is given entirely by (\ref{3.8}). The kinetic
contribution to the shear viscosity, $\eta ^{k}$, is defined by
\begin{equation}
\eta ^{k}=\sum_{i=1}^{s}\;\eta _{i}^{k}=-\frac{1}{(d-1)(d+2)} \sum_{i=1}^{s}\int \,d\mathbf{v}\;m_{i}V_{\alpha
}V_{\beta }\mathcal{C} _{i,\alpha \beta }(\mathbf{V}).  \label{3.2}
\end{equation}
The second equality identifies the partial contribution $\eta _{i}^{k}$ of the species $i$ to the shear
viscosity $\eta ^{k}$ in terms of $\mathcal{C} _{i,\alpha \beta }(\mathbf{V})$, which is the solution to the
integral equation
\begin{equation}
\left( \left( \mathcal{L}+\frac{1}{2}\zeta ^{(0)}\right) \mathcal{C}_{\alpha
\beta }\right) _{i}=C_{i,\alpha \beta },  \label{3.2.1}
\end{equation}
where $C_{i,\alpha \beta }$ is given by Eq.\ (6.18) in the preceding paper \cite {DGH06}. It is symmetric and
traceless so the leading Sonine approximation for the function $\mathcal{C}_{i,\alpha \beta }(\mathbf{V})$ is
\begin{equation}
\mathcal{C}_{i,\alpha \beta }(\mathbf{V})\rightarrow -f_{i,M}(\mathbf{V}) \frac{\eta
_{i}^{k}}{n_{i}T_{i}^{2}}R_{i,\alpha \beta }(\mathbf{V}), \label{3.4bis}
\end{equation}
where
\begin{equation}
R_{i,\alpha \beta }(\mathbf{V})=m_{i}\left( V_{\alpha }V_{\beta }-\frac{1}{d} V^{2}\delta _{\alpha \beta
}\right) .  \label{3.3}
\end{equation}
The partial contributions $\eta _{i}^{k}$ are obtained by multiplying the
integral equation (\ref{3.2.1}) with $R_{i,\alpha \beta }$ and integrating
over the velocity to get the set of equations:
\begin{equation}
\sum_{j=1}^{s}(\tau _{ij}-\frac{1}{2}\zeta ^{(0)}\delta _{ij})\eta
_{j}^{k}=n_{i}T_{i}-\frac{1}{(d-1)(d+2)}\sum_{j=1}^{s}\int d\mathbf{v} \,R_{i,\alpha \beta
}\mathcal{K}_{ij,\alpha }[\partial _{V_{\beta }}f_{j}^{(0)}],  \label{3.4}
\end{equation}
The collision frequencies $\tau _{ii}$ are
\begin{equation}
\tau _{ii}=-\frac{1}{(d-1)(d+2)}\frac{1}{n_{i}T_{i}^{2}}\left( \sum_{j=1}^{s}\int d\mathbf{v}\,R_{i,\alpha \beta
}J_{ij}^{(0)}[f_{i,M}R_{i,\alpha \beta },f_{j}^{(0)}]+\int d\mathbf{v} \,R_{i,\alpha \beta
}J_{ii}^{(0)}[f_{i}^{(0)},f_{i,M}R_{i,\alpha \beta }]\right) ,  \label{3.5}
\end{equation}
\begin{equation}
\tau _{ij}=-\frac{1}{(d-1)(d+2)}\frac{1}{n_{j}T_{j}^{2}}\int d\mathbf{v} \,R_{i,\alpha \beta
}J_{ij}^{(0)}[f_{i}^{(0)},f_{j,M}R_{j,\alpha \beta }],\quad (i\neq j).  \label{3.6}
\end{equation}
The explicit forms of $\tau _{ii}$ and $\tau _{ij}$ are displayed in
Appendix \ref{appA}, and the integral appearing on the right hand side of
Eq.\ (\ref{3.4}) has been evaluated in Appendix \ref{appB} with the result
\begin{eqnarray}
\int d\mathbf{v}\,R_{i,\alpha \beta }\mathcal{K}_{ij,\alpha }[\partial _{V_{\beta }}f_{j}^{(0)}] &=&-\frac{\pi
^{d/2}}{\Gamma \left( \frac{d}{2} \right) }\frac{d-1}{d}m_{i}n_{i}n_{j}\mu _{ji}\sigma _{ij}^{d}\chi
_{ij}^{(0)}(1+\alpha _{ij})  \notag \\
&&\times \left[ \mu _{ji}(3\alpha _{ij}-1)\left( \frac{T_{i}}{m_{i}}+\frac{ T_{j}}{m_{j}}\right)
-4\frac{T_{i}-T_{j}}{m_{i}+m_{j}}\right] .  \label{3.6a}
\end{eqnarray}
In the case of a three-dimensional system ($d=3$), Eq.\ (\ref{3.6a}) reduces
to the one previously derived for hard spheres \cite{GM03}. In addition, for
identical particles previous results \cite{Garzo99,Lutsko05} obtained for
monodisperse gases are also recovered.

With the right side of Eqs.\ (\ref{3.4}) now determined the algebraic
equations
\begin{eqnarray}
\sum_{j=1}^{s}(\tau _{ij}-\frac{1}{2}\zeta ^{(0)}\delta _{ij})\eta _{j}^{k}
&=&n_{i}T_{i}+\frac{m_{i}n_{i}n_{j}\mu _{ji}}{d(d+2)\Gamma \left( \frac{d}{2} \right) }\pi ^{d/2}\sigma
_{ij}^{d}\chi _{ij}^{(0)}(1+\alpha _{ij})  \notag
\\
&&\times \left[ \mu _{ji}(3\alpha _{ij}-1)\left( \frac{T_{i}}{m_{i}}+\frac{ T_{j}}{m_{j}}\right)
-4\frac{T_{i}-T_{j}}{m_{i}+m_{j}}\right] ,  \label{3.7}
\end{eqnarray}
can be solved to determine the partial contributions $\eta _{i}^{k}$. Their sum then gives the kinetic
contribution to the shear viscosity, $\eta ^{k}.$ Finally, adding this to the collisional transfer contribution
of\ (\ref{3.9} ) gives the total shear viscosity.

\subsection{Heat Flux}
\label{sec4}

The constitutive equation for the heat flux $\mathbf{q}^{(1)}$ has
contributions proportional to gradients of the densities and the
temperature, and terms proportional to an applied force (taken to have the
same order of magnitude as the gradients)
\begin{equation}
\mathbf{q}^{(1)}=-\sum_{i=1}^{s}\sum_{j=1}^{s}\left( T^{2}D_{q,ij}\nabla \ln n_{j}+L_{ij}\mathbf{F}_{j}\right)
-T\lambda \nabla \ln T,  \label{4.1}
\end{equation}
where $\lambda$ is the thermal conductivity coefficient and $D_{q,ij}$ are the Dufour coefficients. As in the
case of the shear viscosity, the transport coefficients $D_{q,ij}$ , $L_{ij}$ and $\lambda $ have kinetic and
collisional contributions
\begin{equation}
D_{q,ij}=D_{q,ij}^{k}+D_{q,ij}^{c},\hspace{0.25in} L_{ij}=L_{ij}^{k}+L_{ij}^{c},\hspace{0.25in}\lambda =\lambda
^{k}+\lambda ^{c}.  \label{4.1a}
\end{equation}

The collisional transfer contributions $\lambda ^{c}$, $D_{q,ij}^{c}$, and $ L_{ij}^{c}$ are given in the
preceding paper by Eqs. (7.14)--(7.16)
\begin{eqnarray}
\lambda ^{c} &=&\sum_{i=1}^{s}\sum_{j=1}^{s}\frac{1}{8}\left( 1+\alpha _{ij}\right) m_{j}\mu _{ij}\sigma
_{ij}^{d}\chi _{ij}^{(0)}\left\{ 2B_{4}\left( 1-\alpha _{ij}\right) \left( \mu _{ij}-\mu _{ji}\right) n_{i}
\left[ \frac{2}{m_{j}}\lambda _{j}^{k}+(d+2)\frac{T_{i}}{m_{i}m_{j}T}
\rho D_{j}^{T}\right] \right.  \notag \\
&&\left. +\frac{8B_{2}}{2+d}n_{i}\left[ \frac{2\mu _{ij}}{m_{j}}\lambda
_{j}^{k}-(d+2)\frac{T_{i}}{m_{i}m_{j}T}\left( 2\mu _{ij}-\mu _{ji}\right) \rho D_{j}^{T}\right]
-T^{-1}C_{ij}^{T}\right\} ,  \label{4.1b}
\end{eqnarray}
\begin{eqnarray}
D_{q,ij}^{c} &=&\sum_{p=1}^{s}\frac{1}{8}\left( 1+\alpha _{ip}\right)
m_{p}\mu _{ip}\sigma _{ip}^{d}\chi _{ip}^{(0)}\left\{ 2B_{4}\left( 1-\alpha
_{ip}\right) \left( \mu _{ip}-\mu _{pi}\right) \right.  \notag \\
&&\times n_{i}\left[ \frac{2}{m_{p}}D_{q,pj}^{k}+(d+2)\frac{T_{i}}{
T^{2}}\frac{m_{j}n_{j}}{\rho m_{i}}D_{pj}\right]  \notag \\
&&\left. +\frac{8B_{2}}{d+2}n_{i}\left[ \frac{2\mu _{pi}}{m_{p}} D_{q,pj}^{k}-(d+2)\left( 2\mu _{ip}-\mu
_{pi}\right) \frac{T_{i}}{T^{2} }\frac{n_{j}m_{j}}{\rho m_{i}}D_{pj}\right] -T^{-2}C_{ipj}^{T}\right\} ,
\notag \\
&&  \label{4.1c}
\end{eqnarray}
\begin{eqnarray}
L_{ij}^{c} &=&\sum_{p=1}^{s}\frac{1}{8}\left( 1+\alpha _{ip}\right) m_{p}\mu
_{ip}\sigma _{ip}^{d}\chi _{ip}^{(0)}\left\{ 2B_{4}\left( 1-\alpha
_{ip}\right) \left( \mu _{ip}-\mu _{pi}\right) \right.  \notag \\
&&\times n_{i}\left[ \frac{2}{m_{p}}L_{pj}^{k}+(d+2)\frac{T_{i}}{
m_{i}m_{p}}D_{pj}^{F}\right]  \notag \\
&&\left. +\frac{8B_{2}}{d+2}n_{i}\left[ \frac{2\mu _{pi}}{m_{p}} L_{pj}^{k}-(d+2)\left( 2\mu _{ip}-\mu
_{pi}\right) \frac{T_{i}}{ m_{i}m_{p}}D_{pj}^{F}\right] \right\} .  \label{4.1d}
\end{eqnarray}
The constants, $B_{k}$, are defined by Eq.\ (\ref{b6}) of Appendix \ref{appA}, and
\begin{eqnarray}
C_{ij}^{T} &=&-\frac{2B_{3}}{d}\int d\mathbf{v}_{1}\int d\mathbf{v}
_{2}f_{i}^{(0)}(\mathbf{V}_{1})f_{j}^{(0)}(\mathbf{V}_{2})\left\{ gG_{ij}^{2}+g^{-1}\left( \mathbf{g}\cdot
\mathbf{G}_{ij}\right) ^{2}+(1+\mu
_{ji})g\left( \mathbf{g}\cdot \mathbf{G}_{ij}\right) \right.  \notag \\
&&\left. +\mu _{ji}\mu _{ij}g^{3}+\frac{3}{4}(1-\alpha _{ij})(\mu _{ji}-\mu
_{ij})\left[ g\left( \mathbf{g}\cdot \mathbf{G}_{ij}\right) +g^{3}\right]
\right\} ,  \label{4.1e}
\end{eqnarray}
\begin{equation}
C_{ipj}^{T}=\frac{B_{3}}{d}\int d\mathbf{v}_{1}\int d\mathbf{v}_{2}\left[ -\left( 1-\alpha _{ip}\right) \left(
\mu _{ip}-\mu _{pi}\right) g^{3}+4g\left( \mathbf{g}\cdot \mathbf{G}_{ip}\right) \right] f_{i}^{(0)}(
\mathbf{V}_{1})n_{j}\partial _{n_{j}}f_{p}^{(0)}(\mathbf{V}_{2}), \label{4.1f}
\end{equation}
where $\mathbf{G}_{ip}=\mu _{ip}\mathbf{V}_{1}+\mu _{pi}\mathbf{V}_{2}$. In the first Sonine approximation these
integrals have the explicit forms
\begin{eqnarray}
C_{ij}^{T} &=&-\frac{2\pi ^{(d-1)/2}}{d\Gamma \left( \frac{d}{2}\right) } n_{i}n_{j}v_{0}^{3}(\theta _{i}+\theta
_{j})^{-1/2}(\theta _{i}\theta
_{j})^{-3/2}  \notag \\
&&\times \left\{ 2\beta _{ij}^{2}+\theta _{i}\theta _{j}+(\theta _{i}+\theta
_{j})\left[ (\theta _{i}+\theta _{j})\mu _{ij}\mu _{ji}+\beta _{ij}(1+\mu
_{ji})\right] \right\}  \notag \\
&&-\frac{3\pi ^{(d-1)/2}}{2d\Gamma \left( \frac{d}{2}\right) } n_{i}n_{j}v_{0}^{3}(1-\alpha _{ij})(\mu _{ji}-\mu
_{ij})\left( \frac{\theta _{i}+\theta _{j}}{\theta _{i}\theta _{j}}\right) ^{3/2}\left[ \mu
_{ji}+\beta _{ij}(\theta _{i}+\theta _{j})^{-1}\right] ,  \notag \\
&&  \label{4.1g}
\end{eqnarray}
\begin{eqnarray}
C_{ipj}^T&=&\frac{4\pi^{(d-1)/2}}{d\Gamma\left(\frac{d}{2}\right)}n_in_pv_0^3
(\theta_i+\theta_p)^{-1/2}(\theta_i\theta_p)^{-3/2}\left\{
\delta_{jp}\beta_{ip}(\theta_i+\theta_p)\right.\nonumber\\
& & \left. -\frac{1}{2}\theta_i\theta_p \left[1+\frac{\mu_{pi}(\theta_i+\theta_p)-2 \beta_{ip}}{\theta_p}\right]
\frac{\partial \ln \gamma_p}{\partial \ln
n_j}\right\}\nonumber\\
& & +\frac{\pi^{(d-1)/2}}{d\Gamma\left(\frac{d}{2}\right)}n_in_pv_0^3 (1-\alpha_{ip})(\mu_{pi}-\mu_{ip})
\left(\frac{\theta_i+\theta_p}{\theta_i\theta_p}\right)^{3/2} \left( \delta_{jp}
+\frac{3}{2}\frac{\theta_i}{\theta_i+\theta_p}\frac{\partial \ln \gamma_p}{\partial \ln
n_j}\right),\nonumber\\
\label{4.1h}
\end{eqnarray}
where $\beta _{ip}\equiv \mu _{ip}\theta _{p}-\mu _{pi}\theta _{i}$. With these expressions, the collisional
contributions to the heat flux are explicitly known. For mechanically equivalent particles, all the above
expressions for the collision transfer contributions reduce again to the previous results obtained for
monocomponent gases \cite{Garzo99a,Lutsko05}.

\subsubsection{Kinetic contributions}
\label{subsec4.1}

The kinetic parts of the transport coefficients $\lambda ^{k}$, $ D_{q,ij}^{k} $, $L_{ij}^{k}$ are defined
respectively as
\begin{equation}
\lambda ^{k}=-\frac{1}{dT}\sum_{i=1}^{s}\;\int d\mathbf{v}\,\frac{m_{i}}{2} V^{2}\mathbf{V}\cdot
\boldsymbol{\mathcal{A}}_{i}(\mathbf{V}),  \label{4.2}
\end{equation}
\begin{equation}
D_{q,ij}^{k}=-\frac{1}{dT^{2}}\int d\mathbf{v}\,\frac{m_{i}}{2}V^{2}\mathbf{V }\cdot
\boldsymbol{\mathcal{B}}_{i}^{j}(\mathbf{V}),  \label{4.2bis}
\end{equation}
\begin{equation}
L_{ij}^{k}=-\frac{1}{d}\int d\mathbf{v}\,\frac{m_{i}}{2}V^{2}\mathbf{V}\cdot
\boldsymbol{\mathcal{E}}_{i}^{j}(\mathbf{V}).  \label{4.2.1}
\end{equation}
where $\boldsymbol{\mathcal{A}}_{i}(\mathbf{V})$, $\boldsymbol{\mathcal{B}} _{i}^{j}(\mathbf{V})$, and
$\boldsymbol{\mathcal{E}}_{i}^{j}(\mathbf{V})$ are again the solutions to (\ref{2.3.2})--(\ref{2.3.4}). The
leading Sonine approximations of (\ref{2.4})--(\ref{2.5.1}) are not adequate to determine the general leading
order for these transport coefficients, since the coefficients $D_{i}^{T},$ $D_{ij}$, and $D_{ij}^{F}$ vanish
for a simple one component fluid. This would imply a vanishing thermal conductivity $\lambda$ as well.
Consequently, it is necessary to include here the next (second order) Sonine polynomial
\begin{equation}
\boldsymbol{\mathcal{A}}_{i}(\mathbf{V})\rightarrow f_{i,M}(\mathbf{V})\left[ -\frac{\rho
}{n_{i}T_{i}}\mathbf{V}D_{i}^{T}-\frac{2}{d+2}\frac{Tm_{i}}{ n_{i}T_{i}^{3}}\lambda
_{i}\mathbf{S}_{i}(\mathbf{V})\right] ,  \label{4.3}
\end{equation}
\begin{equation}
\boldsymbol{\mathcal{B}}_{i}^{j}(\mathbf{V})\rightarrow f_{i,M}(\mathbf{V}) \left[ -\frac{m_{i}m_{j}n_{j}}{\rho
n_{i}T_{i}}\mathbf{V}D_{ij}-\frac{2}{d+2}
\frac{T^{2}m_{i}}{n_{i}T_{i}^{3}}d_{q,ij}\mathbf{S}_{i}(\mathbf{V})\right] , \label{4.4}
\end{equation}
\begin{equation}
\boldsymbol{\mathcal{E}}_{i}^{j}(\mathbf{V})\rightarrow f_{i,M}(\mathbf{V}) \left[
-\frac{1}{n_{i}T_{i}}\mathbf{V}D_{ij}^{F}-\frac{2}{d+2}\frac{m_{i}}{ n_{i}T_{i}^{3}}\ell
_{ij}\mathbf{S}_{i}(\mathbf{V})\right] ,  \label{4.4.1}
\end{equation}
where the next order polynomial $\mathbf{S}_{i}(\mathbf{V})$ is
\begin{equation}
\mathbf{S}_{i}(\mathbf{V})=\left( \frac{1}{2}m_{i}V^{2}-\frac{d+2}{2} T_{i}\right) \mathbf{V}.  \label{4.5}
\end{equation}
In the above equations, it is understood that the transport coefficients $ D_{i}^{T}$, $D_{ij}$, and
$D_{ij}^{F}$ are given by Eqs.\ (\ref{2.13})--( \ref{2.14.1}), respectively. The coefficients $\lambda _{i}$,
$d_{q,ij}$, and $\ell _{ij}$ are the projections along $\mathbf{S}_{i}$:
\begin{equation}
\lambda _{i}=-\frac{1}{dT}\int \;d\mathbf{v}\;\mathbf{S}_{i}(\mathbf{V} )\cdot
\boldsymbol{\mathcal{A}}_{i}(\mathbf{V}),  \label{4.6}
\end{equation}
\begin{equation}
d_{q,ij}=-\frac{1}{dT^{2}}\int \;d\mathbf{v}\;\mathbf{S}_{i}(\mathbf{V} )\cdot
\boldsymbol{\mathcal{B}}_{i}^{j}(\mathbf{V}),  \label{4.7}
\end{equation}
\begin{equation}
\ell _{ij}=-\frac{1}{d}\int \;d\mathbf{v}\;\mathbf{S}_{i}(\mathbf{V})\cdot
\boldsymbol{\mathcal{E}}_{i}^{j}(\mathbf{V}).  \label{4.7.1}
\end{equation}
In terms of $\lambda _{i}$, $d_{q,ij}$ and $\ell _{ij}$, the transport
coefficients $\lambda ^{k}$, $D_{q,ij}^{k}$, and $L_{ij}^{k}$ become
\begin{equation}
\lambda ^{k}=\sum_{i=1}^{s}\;\lambda _{i}+\frac{d+2}{2T}\frac{\rho T_{i}}{ m_{i}}D_{i}^{T},  \label{4.8}
\end{equation}
\begin{equation}
D_{q,ij}^{k}=d_{q,ij}+\frac{d+2}{2T^{2}}\frac{m_{j}n_{j}T_{i}}{\rho }D_{ij}.
\label{4.9}
\end{equation}
\begin{equation}
L_{ij}^{k}=\ell _{ij}+\frac{d+2}{2}\frac{T_{i}}{m_i}D_{ij}^{F}.  \label{4.9.1}
\end{equation}

Since $D_{i}^{T}$, $D_{ij}$, and $D_{ij}^{F}$ are known from the analysis above it remains to determine $\lambda
_{i}$, $d_{q,ij}$ and $\ell _{ij}$. The algebraic equations determining them are obtained by substituting (\ref
{4.3})--(\ref{4.4.1}) into the integral equations (\ref{2.3.2})--(\ref {2.3.4}), multiplying Eq.\ (\ref{2.3.2})
by $\mathbf{S}_{i}(\mathbf{V})$ and integrating over the velocity. The results are
\begin{equation}
\sum_{j=1}^{s}(\gamma _{ij}-2\zeta ^{(0)}\delta _{ij})\lambda _{j}=\overline{ \lambda }_{i},  \label{4.10}
\end{equation}
\begin{equation}
\sum_{\ell =1}^{s}\left( \gamma _{i\ell }-\frac{3}{2}\zeta ^{(0)}\delta
_{i\ell }\right) d_{q,\ell j}=\overline{d}_{q,ij}  \label{4.11}
\end{equation}
\begin{equation}
\sum_{k=1}^{s}\left( \gamma _{ik}-\frac{1}{2}\zeta ^{(0)}\delta _{ik}\right)
\ell _{kj}=\overline{\ell }_{ij},  \label{4.12}
\end{equation}
with the inhomogeneities given by
\begin{eqnarray}
\overline{\lambda }_{i} &=&-\frac{d+2}{2}\frac{\rho }{T_{i}}\frac{
n_{i}T_{i}^{3}}{Tm_{i}}\sum_{j=1}^{s}\frac{\omega _{ij}-\zeta ^{(0)}\delta
_{ij}}{n_{j}T_{j}}D_{j}^{T}+\frac{d(d+2)}{2d}\frac{n_{i}T_{i}^{2}}{m_{i}T}
\notag \\
&&-\frac{1}{2Td}\sum_{j=1}^{s}\int d\mathbf{v}\,S_{i,\beta }(\mathbf{V}) \mathcal{K}_{ij,\beta }[\nabla
_{\mathbf{V}}\cdot (\mathbf{V}f_{j}^{(0)})], \label{4.13}
\end{eqnarray}
\begin{eqnarray}
\overline{d}_{q,ij} &=&-\frac{d+2}{2}\frac{n_{i}n_{j}T_{i}^{3}}{m_{i}T^{2}} \left( \frac{m_{j}}{\rho
T_{i}}\sum_{\ell =1}^{s}m_\ell\frac{\omega _{i\ell }-\zeta ^{(0)}\delta _{i\ell }}{n_{\ell }T_{\ell }}D_{\ell
j}+\frac{\partial \zeta ^{(0)}}{\partial n_{j}}\lambda _{i}-\frac{1}{T_{i}}\frac{\partial \ln
T_{i}}{\partial n_{j}}\right)  \notag \\
&&+\frac{1}{dT^2}\sum_{\ell =1}^{s}\int d\mathbf{v}\,S_{i,\gamma }\left( \mathcal{K}_{i\ell ,\gamma
}[n_{j}\partial _{n_{j}}f_{\ell }^{(0)}]+\frac{1}{ 2}\left( n_{\ell }\partial _{n_{j}}\ln \chi _{i\ell
}^{(0)}+I_{i\ell j}\right) \mathcal{K}_{i\ell ,\gamma }[f_{\ell }^{(0)}]\right),\nonumber\\
\label{4.13a}
\end{eqnarray}
\begin{equation}
\overline{\ell}_{ij}=-\frac{d+2}{2}\frac{n_{i}T_{i}^{2}}{m_{i}} \sum_{k=1}^{s}\frac{\omega _{ik}-\zeta
^{(0)}\delta _{ik}}{n_{k}T_{k}} D_{kj}^{F}.  \label{4.13b}
\end{equation}
Use has been made of the results
\begin{equation}
\int d\mathbf{v}\,\mathbf{S}_{i}\cdot \mathbf{A}_{i}({\bf V})=-\frac{d(d+2)}{2}\frac{
n_{i}T_{i}^{2}}{m_{i}}+\frac{1}{2}\sum_{j=1}^{s}\int d\mathbf{v}\,S_{i,\beta }(\mathbf{V})\mathcal{K}_{ij,\beta
}[\nabla _{\mathbf{V}}\cdot (\mathbf{V} f_{j}^{(0)})],  \label{4.15}
\end{equation}
\begin{eqnarray}
\int d\mathbf{v}\,\mathbf{S}_{i}\cdot \mathbf{B}_{i}^{j}(\mathbf{ V})&=&-\frac{d(d+2)}{2}\frac{
n_{i}n_jT_{i}}{m_{i}}\frac{\partial T_i}{\partial n_j}\nonumber\\
& &  -\sum_{\ell=1}^{s}\int d\mathbf{v}\,S_{i,\beta }(\mathbf{V})\mathcal{K}_{i\ell,\beta}\left[\left(
n_{j}\partial _{n_{j}}+\frac{1}{2}\left( n_{\ell }\frac{\partial \ln \chi _{i\ell }^{(0)}}{\partial
n_{j}}+I_{i\ell j}\right) \right) f_{\ell }^{(0)} \right], \label{4.15.1}
\end{eqnarray}
\begin{equation}
\int d\mathbf{v}\,\mathbf{S}_{i}(\mathbf{V})\cdot \mathbf{E}_{i}^{j}(\mathbf{ V})=0.  \label{4.17.2}
\end{equation}
The collision frequencies introduced \ here are
\begin{equation}
\gamma _{ii}=-\frac{2}{d(d+2)}\frac{m_{i}}{n_{i}T_{i}^{3}}\left( \sum_{j=1}^{s}\int
d\mathbf{v}\,\mathbf{S}_{i}\cdot J_{ij}^{(0)}[f_{i,M} \mathbf{S}_{i},f_{j}^{(0)}]+\int
d\mathbf{v}\,\mathbf{S}_{i}\cdot J_{ii}^{(0)}[f_{i}^{(0)},f_{i,M}\mathbf{S}_{i}]\right) ,  \label{4.13c}
\end{equation}
\begin{equation}
\gamma _{ij}=-\frac{2}{d(d+2)}\frac{m_{j}}{n_{j}T_{j}^{3}}\int d\mathbf{v}\, \mathbf{S}_{i}\cdot
J_{ij}^{(0)}[f_{i}^{(0)},f_{j,M}\mathbf{S}_{j}],\quad (i\neq j).  \label{4.13d}
\end{equation}
\begin{equation}
\omega _{ii}=-\frac{2}{d(d+2)}\frac{m_{i}}{n_{i}T_{i}^{2}}\left( \sum_{j=1}^{s}\int
d\mathbf{v}\,\mathbf{S}_{i}\cdot J_{ij}^{(0)}[f_{i,M} \mathbf{V},f_{j}^{(0)}]+\int
d\mathbf{v}\,\mathbf{S}_{i}\cdot J_{ii}^{(0)}[f_{i}^{(0)},f_{i,M}\mathbf{V}]\right) ,  \label{4.13e}
\end{equation}
\begin{equation}
\omega _{ij}=-\frac{2}{d(d+2)}\frac{m_{i}}{n_{i}T_{i}^{2}}\sum_{j=1}^{s}\int d\mathbf{v}\,\mathbf{S}_{i}\cdot
J_{ij}^{(0)}[f_{i}^{(0)},f_{j,M}\mathbf{V} ].\quad (i\neq j).  \label{4.13f}
\end{equation}
The explicit expressions for the above collision frequencies are also displayed in Appendix \ref{appA}.

The inhomogeneous terms (\ref{4.13}) and (\ref{4.13a}) involve integrals over the operator
$\mathcal{K}_{ij,\gamma }$. These are evaluated in Appendix \ref{appB}
\begin{eqnarray}
\int d\mathbf{v} &&S_{i,\gamma }(\mathbf{V})\mathcal{K}_{ij,\gamma }[\nabla
_{\mathbf{V}}\cdot (\mathbf{V}f_{j}^{(0)})]=-\frac{\pi ^{d/2}}{\Gamma \left(
\frac{d}{2}\right) }n_{i}n_{j}\mu _{ij}\chi _{ij}^{(0)}\sigma
_{ij}^{d}T_{j}(1+\alpha _{ij})\left\{ \frac{T_{i}}{m_{i}}\left[ (d+2)(\mu
_{ij}^{2}-1)\right. \right.  \notag \\
&&\left. \left. +(2d-5-9\alpha _{ij})\mu _{ij}\mu _{ji}+(d-1+3\alpha _{ij}+6\alpha _{ij}^{2})\mu
_{ji}^{2}\right] +6\frac{T_{j}}{m_{j}}\mu _{ji}^{2}(1+\alpha _{ij})^{2}\right\} ,  \label{4.15a}
\end{eqnarray}
\begin{eqnarray}
\int d\mathbf{v} &&S_{i,\gamma }(\mathbf{V})\mathcal{K}_{ij,\gamma }[f_{j}^{(0)}]=\frac{\pi ^{d/2}}{2\Gamma
\left( \frac{d}{2}\right) } m_{i}n_{i}n_{j}\chi _{ij}^{(0)}\sigma _{ij}^{d}\mu _{ji}(1+\alpha _{ij})
\notag \\
&&\times \left\{ \left[ (d+8)\mu _{ij}^{2}+(7+2d-9\alpha _{ij})\mu _{ij}\mu _{ji}+(2+d+3\alpha _{ij}^{2}-3\alpha
_{ij})\mu _{ji}^{2}\right] \frac{
T_{i}^{2}}{m_{i}^{2}}\right.  \notag \\
&&+3\mu _{ji}^{2}(1+\alpha _{ij})^{2}\frac{T_{j}^{2}}{m_{j}^{2}}+\left[
(d+2)\mu _{ij}^{2}+(2d-5-9\alpha _{ij})\mu _{ij}\mu _{ji}+(d-1+3\alpha
_{ij}+6\alpha _{ij}^{2})\mu _{ji}^{2}\right]  \notag \\
&&\left. \times \frac{T_{i}T_{j}}{m_{i}m_{j}}-(d+2)\left( \frac{T_{i}}{m_{i}} +\frac{T_{j}}{m_{j}}\right)
\frac{T_{i}}{m_{i}}\right\} ,  \label{4.17}
\end{eqnarray}
\begin{equation}
\int d\mathbf{v}\,S_{i,\gamma }(\mathbf{V})\mathcal{K}_{i\ell ,\gamma }[n_{j}\partial _{n_{j}}f_{\ell
}^{(0)}]=\int d\mathbf{v}\,S_{i,\gamma }( \mathbf{V})\left( \mathcal{K}_{i\ell ,\gamma }[\delta _{\ell j}f_{\ell
}^{(0)}]-\frac{1}{2}\mathcal{K}_{i\ell ,\gamma }[\nabla _{\mathbf{V}}\cdot ( \mathbf{V}f_{\ell
}^{(0)})]\frac{\partial \ln \gamma _{\ell }}{\partial \ln n_{j}}\right).  \label{4.18}
\end{equation}
This completely determines the parameters of the integral equations (\ref {4.10})--(\ref{4.12}). Their solution
determines $\lambda _{i}$, $d_{q,ij}$ and $\ell _{ij}$, and hence the transport coefficients $\lambda ^{k}$, $
D_{q,ij}^{k}$, and $L_{ij}^{k}$ through Eqs.\ (\ref{4.8})--(\ref{4.9.1}). In the special case of mechanically
equivalent particles, this description reduces to those previously obtained for a monodisperse gas
\cite{Garzo99a,Lutsko05}.

\section{Summary and Discussion}
\label{sec6}

First, a brief summary of the results obtained in the two papers presented here is given. The revised Enskog
kinetic equation was used as the basis for deriving exact balance equations for the hydrodynamic fields. The
pressure, cooling rate, mass flux, momentum flux, and energy flux were determined from the kinetic theory as
linear combinations of the gradients of these fields and an applied force. These are the constitutive equations
that convert the exact balance equations into closed equations for the hydrodynamic fields (Navier-Stokes
equations, at this order of gradients). The coefficients in these equations are the pressure, leading order
cooling rate, and the transport coefficients. Formally exact expressions for these coefficients were obtained in
the first paper, expressed as integrals over solutions to integral equations via the Chapman-Enskog method for
constructing a solution to the kinetic equation. In the present paper, these exact expressions were evaluated
using approximate solutions to the integral equations. The approximation consists of expanding the unknown
functions in a complete set of polynomials, and truncating that expansion to convert the integral equations to
algebraic equations that can be solved by standard matrix methods. Here, the detailed forms for those equations
or their solutions have been given in terms of the hydrodynamic fields and other parameters of the problem. The
result is a complete description of the hydrodynamic equations to Navier-Stokes order with all parameters
determined explicitly from the theory. Due to the complexity of the analysis it may be useful to give the
specific location of the final results. The pressure is given by Eq. (\ref{1.2}) and the lowest order cooling
rate by Eq.\ (\ref{1.3}). The equivalence of all species cooling rates determine the species temperatures as
well (to this order in the gradients). The only transport coefficient for the cooling rate to first order is
given by (\ref{1.17}) and the solution to the algebraic equations (\ref{1.13}). The transport coefficients
characterizing the mass flux are the solutions to the algebraic equations (\ref{2.13})--(\ref{2.14.1}). The
momentum flux has two transport coefficients, the shear and bulk viscosities. These are given by Eqs.\
(\ref{3.1a})--(\ref{3.2}) and the solutions to the algebraic equations (\ref{3.7}). Finally, the transport
coefficients for the heat flux are given by Eqs.\ (\ref{4.1a})--(\ref{4.1d}), (\ref{4.8})--(\ref {4.9.1}), and
the solutions to the algebraic equations (\ref{4.10})--(\ref {4.12}). Detailed forms for the collision
frequencies and other input functions are presented in the Appendices. These explicit ``constitutive relations''
together with the exact macroscopic balance equations for species mass densities, flow velocity, and temperature
(Eqs.\ (4.12), (4.14), and (4.24) of the previous paper) complete the practical description of Navier-Stokes
hydrodynamics for a moderately dense, multi-component granular mixture based on the revised Enskog kinetic
equation.

In the remainder of this Discussion the relationship to several previous works on kinetic-theory-based
descriptions of granular mixtures in the literature is considered. These contributions, which are abbreviated
according to the initials of authors and the final two digits of the publication year, are listed in Tables
\ref{CMH.Tab.1} and \ref{CMH.Tab.2}. To help put the current effort in the context of previous works, Tables
\ref {CMH.Tab.1} and \ref{CMH.Tab.2} also list the applicability of each contribution (dimensionality and number
of species), the starting kinetic equation (Boltzmann or Enskog), and the specific mechanics and assumptions
used in the derivation process: the standard Enskog theory (SET) vs.\ the revised Enskog theory (RET), pair
correlation function $\chi_{ij}^{(0)}$, solution method, order and base state of expansion, single particle
velocity distribution function [Maxwellian (M) and non-Maxwellian (nM)], energy distribution [energy
equipartition (EE) and non-energy equipartition (nEE)], and hydrodynamic variables. In the subsections below,
each of these various treatments are detailed and their implications are discussed.\bigskip

\noindent\textit{Dimensionality}
\bigskip

The number contained in Table \ref{CMH.Tab.1} refers to the dimensionality of the particles. Namely, 2D refers
to circular particles (disks) that are constrained to motion in a plane, while 3D refers to spherical particles
that can move in all three dimensions. Although systems of practical importance are three-dimensional,
two-dimensional theories have appeared in the literature for purposes of comparing with molecular-dynamics (MD)
simulations. Specifically, early MD simulations were performed in two dimensions due to computational
constraints, whereas three-dimensional simulations are now common. The WA99 theory is applicable to 2D only, the
JM87 and GHD theories are derived for both 2D and 3D, while all remaining theories JM89, Z95, H06, GD02, R03,
IA05, and S06 are for 3D systems.
\bigskip

\noindent\textit{Number of Species}
\bigskip

Most mixture theories to date have been targeted at binary mixtures, in which the two species can differ in
size, mass, restitution coefficient, and density (JM87, JM89, WA99, H01, GD02, R03, and S06). Some recent
theories have been derived for a more general system of $s$ distinct species (Z95, IA05, and GHD).
\bigskip

\noindent\textit{Kinetic Theory}
\bigskip

Two kinetic theories have been used in the development of hydrodynamic equations for mixtures, namely the
Boltzmann equation and the Enskog equation for hard spheres. The difference between these two equations stems
from the treatment of the two-particle distribution function $f_{ij}$. For the Boltzmann equation, $f_{ij}$ is
assumed equal to the product of the two single-particle distribution functions $(f_{ij}=f_{i}f_{j})$. This lack
of spatial and pre-collisional velocity correlations between the two particles restricts the Boltzmann equation
to dilute systems. The Enskog equation, on the other hand, accounts for positional correlations (but not
pre-collisional velocity correlations) via the equilibrium pair correlation function at contact $\chi_{ij}$,
namely $f_{ij}=\chi _{ij}f_{i}f_{j}$ (the Enskog approximation). More specifically, $\chi_{ij}$ accounts for
excluded volume effects encountered in denser flows, and thus the corresponding Enskog kinetic theory is
applicable to moderately dense flows. The Enskog approximation is expected to deteriorate at higher densities as
ring collisions and their associated velocity correlations become important. The prevalence of such correlations
has also been found to depend on the restitution coefficient; namely, velocity correlations have been observed
to increase as the restitution coefficient decreases. Thus, theories based on the Enskog equation are applicable
to both dilute and moderately dense flows (JM87, JM89, Z95, WA99, H01, R03, IA05, and GHD), while theories using
the Boltzmann equation as a starting point are restricted to dilute flows (GD02 and S06).

From a practical perspective, the upper limit of concentration that a given kinetic theory should be applied to
depends on the desired level of accuracy. As a quick gauge to the range of validity of the Boltzmann equation,
the limiting case of a monodisperse system is considered using the theories of Garz\'{o} and Dufty
\cite{Garzo99a} and Lutsko \cite{Lutsko05}, which are based on the Enskog equation and thus are applicable to
both dilute and dense flows. According to these theories, the dense collisional contributions to the pressure
(absent in the Boltzmann equation) are 4\% and 18\% for solids volume fractions of 0.01 and 0.05, respectively,
at $\alpha =0.9$. A similar estimate of the range of validity of the Enskog equation is not available, since the
impact of velocity correlations on granular hydrodynamics has not been extensively studied.\bigskip

\noindent\textit{SET vs.\ RET}
\bigskip

For those contributions listed in Table \ref{CMH.Tab.1} that employ the Enskog equation, two different
approaches are possible -- the SET and the RET. As mentioned previously, the difference between SET and RET
traces to the choice of the pair correlation function, $\chi _{ij}^{(0)}$. In SET, $\chi _{ij}^{(0)}$ is a
\textit{ function} of concentration (i.e., depends on local value only) at a single position of interest,
whereas for RET $\chi _{ij}^{(0)}$ is treated as a \textit{ functional} of concentration (i.e., depends on the
local value and its gradient) at the two particle centers. In SET, the location (e.g., midpoint) at which to
evaluate $\chi _{ij}^{(0)}$ for mixtures is unclear \cite{Barajas73}. Regardless of the choice, however, the
resulting diffusion force is found to be inconsistent with irreversible thermodynamics, unlike in RET \cite
{vanBeijeren73}.

The implications of the SET vs. RET treatment on the resulting theory depends on the type system being examined.
For the case of monodisperse systems, SET and RET lead to the same Navier-Stokes transport coefficients
\cite{vanBeijeren73,vanBeijeren73a,vanBeijeren79,LG93} but different inhomogeneous equilibrium states
\cite{vanBeijeren73,vanBeijeren73a,vanBeijeren79}. For mixtures, however, different Navier-Stokes transport
coefficients are obtained from SET and RET. More specifically, although the fluxes appearing in the momentum and
energy balances are the same for SET and RET [${\bf q}$ and $P_{\gamma \beta }$ in Eqs.\ (4.13) and (4.14) in
the companion paper], the diffusion flux [${\bf j}_{0i}$ in Eq.\ (4.12) in companion paper] takes on different
forms \cite{Arnarson00}. Although the quantitative impact of such differences on segregation predictions has not
been investigated in detail, it is clear that RET is the appropriate approach since it is consistent with
irreversible thermodynamics. The Enskog-based theories of JM89, Z95, WA99, and GHD use RET, whereas JM87, H01,
R03, and IA05 utilize SET.
\bigskip

\noindent\textit{Pair Correlation Function}
\bigskip

As noted in Table \ref{CMH.Tab.1}, various forms of the pair correlation function $\chi _{ij}^{(0)}$ have been
used in conjunction with mixtures. Its explicit forms can be found in the references displayed in the
table.\bigskip

\noindent\textit{Solution Method}
\bigskip

With the exception of Zamankhan \cite{Zamankhan95}, all of the efforts to date on mixtures have implemented a
Chapman-Enskog (CE) expansion to solve the kinetic equation. By definition, the CE expansion involves a
perturbative expansion about low Knudsen numbers (where $\mathrm{Kn}$ is defined as the ratio of the mean free
path to the characteristic length of the mean-flow gradients) or ``small gradients,'' and thus is not applicable
to systems in which free-molecular (non-continuum) effects play a non-negligible role. The Grad moment method,
as was employed by Zamankhan \cite{Zamankhan95} does not contain similar restrictions, though the derivation is
necessarily more complex and thus has not been performed without resorting to other simplifying assumptions
(e.g., equipartition of energy).
\bigskip

\noindent\textit{Order of Chapman-Enskog Expansion}
\bigskip

For each of the entries in Table \ref{CMH.Tab.1} that employ the CE expansion to solve the kinetic theory
equation, all expansions are carried out to first order in spatial gradients -- i.e., to the Navier Stokes (NS)
order. Nonetheless, evidence of higher-order effects has been noted in a range of granular flows, and
Burnett-order effects (second order in spatial gradients) in particular have been shown to be linked to the
anisotropy in the stress tensor \cite{Sela98}. Examples of systems in which significant higher-order effects
have been identified include the upper region of an open-ended, vibro-fluidized bed \cite{Brey01, Martin05}, the
vanishing heap of a vibrated, granular material \cite{Behringer02}, dilute flow around an immersed cylinder
\cite{Wassgren03}, simple shear flow \cite{Santos04}, the Knudsen layer adjacent to a thermal (energy-providing)
boundary \cite {Galvin07}, and in the continuum interior of a thermally-driven granular gas \cite{Hrenya06}. For
the case of monodisperse systems, several theories\cite {Kumaran97,Sela98,Risso02,Kumaran05} and boundary
conditions\cite {Goldhirsch99,Brey01,Martin05} have been developed that account for such higher order effects in
a limited class of systems (dilute, sheared flow, open boundary, etc.). Analogous work has not been reported for
mixtures \cite{MG02,L04}, however, and thus the incorporation of higher-order effects does not serve as a
differentiator between those theories listed in Table \ref{CMH.Tab.2} that employ the CE expansion. Nonetheless,
the possible presence of such higher-order effects should be kept in mind when comparing these theories with
experiments and simulations for purposes of validation.
\bigskip

\noindent\textit{Base State of Chapman-Enskog Expansion}
\bigskip

As described in the companion paper, hydrodynamics results from a ``normal solution'' to the kinetic equation,
whose space and time dependence occurs only through the hydrodynamic fields. The CE expansion is a systematic
method for constructing this normal solution as an expansion in powers of the Knudsen number, or spatial
gradients of the fields. At zeroth order in these fields, the kinetic theory determines the form of the
distribution function to be the ``local equilibrium'' Maxwellian for molecular fluids. However, in the presence
of dissipation the kinetic theory requires a different solution at zeroth order, the ``local homogeneous
cooling'' (HCS) distribution. The HCS distribution agrees with the local Maxwellian only for $\alpha =1$. It is
possible to make an expansion in both Knudsen number and $ \left( 1-\alpha \right) $, in which case the lowest
order term is indeed the Maxwellian. Such a double expansion is necessarily limited to asymptotically weak
dissipation. The theories of JM87, JM89, WA99, H01, R03, IA05, and S06 are of this type with the corresponding
implicit limitation. The theories of GD02 and GHD, on the other hand, expand only in the Knudsen number, with
HCS as the leading order solution and hence no a priori limitation on the degree of dissipation. Quantitatively,
the difference between the two types of expansions has been examined in monodisperse systems via a comparison of
the dissipation rate obtained in MD simulations \cite{Hrenya06}. As an example, at a volume fraction of 0.3 and
$\alpha =0.75$, an error of 23\% is obtained when using a theory \cite{Jenkins98} based on an expansion about
$\alpha =1$ , while an error of 7\% is obtained when using a theory \cite{Garzo99a} based on an expansion about
the HCS. For both theories, the level of mismatch is found to increase with concentration and dissipation
levels.
\bigskip

\noindent\textit{Order of Sonine polynomial expansion}
\bigskip

To make the analytical evaluation of the collision integrals possible, a
truncated Sonine polynomial expansion is employed. All theories listed in
Table \ref{CMH.Tab.1} employ the lowest, non-zero order of the polynomial
expansion (leading term), except for the weak dissipation theory of S06
which carries out the expansion to third order. Nonetheless, the transport
coefficients of S06 agree with those of GD02 in the common domain of
validity, namely in the nearly elastic limit \cite{Serero06}.

The accuracy of a given Sonine polynomial approximation can be tested via comparison with discrete simulation
Monte Carlo (DSMC) results. DSMC provides a numerical solution of the starting kinetic (Boltzmann or Enskog)
equation for a specific system, and thus provides a check for the existence of a normal solution, the order of
the gradient expansion (e.g., Navier Stokes), and the truncated Sonine polynomial. For dilute granular mixtures,
good agreement between DSMC and the theory of Garz\'{o} and Dufty \cite {Garzo02} was found for the shear
viscosity, even for strong dissipation \cite {Montanero03}. This agreement for the shear viscosity is also
present for moderately dense systems \cite{GM03}. A similar comparison was carried out for the diffusion
coefficient for a system with an impurity \cite{Garzo04}. The agreement is again excellent, except when the size
or mass disparity is large, in which case the second Sonine approximation leads to a significant improvement
over the first Sonine approximation.
\bigskip

\noindent\textit{Single Particle Velocity Distribution}
\bigskip

In an effort to simplify the evaluation of collision integrals, some previous works have assumed that the single
particle velocity distribution function is Maxwellian (M). Strictly speaking, such an assumption is valid only
for perfectly elastic spheres in equilibrium \cite{Chapman70}. Numerous experimental, theoretical, and
simulation studies of inelastic, monodisperse systems have indicated that the distribution function departs from
Maxwellian \cite{Campbell90, Goldshtein95, Goldhirsch96, Esipov97, vanNoije98, Brey99, Losert99, Kudrolli00}.
For the case of perfectly elastic mixtures, an estimate of the impact of this effect is given by Willits and
Arnarson \cite{Willits99}, who compare the shear viscosity predictions of two theories with that of MD
simulations. For this particular system, the only effective difference between the two theories is that one
contains a Maxwellian assumption and the other does not. For disks with a diameter ratio of 1.25 and over a
range of solids fractions from 0.05 to 0.4, the non-Maxwellian (nM) theory exhibits very good agreement with the
simulations. The Maxwellian-based predictions, on the other hand, are significantly lower in value since the
Maxwellian assumption precludes the kinetic contribution to the shear stress and simplifies its collisional
counterpart. Although similar studies on the impact of non-Maxwellian effects on other constitutive quantities
are not available, it is clear that a non-Maxwellian treatment is critical for the accurate prediction of shear
stress. Of the mixture theories presented in Table \ref{CMH.Tab.1}, those of JM87, H01, and R03 are based on a
Maxwellian assumption, while those of JM89, WA99, Z95, GD02, S06, and GHD incorporate non-Maxwellian effects.
The theory of IA05 takes a hybrid approach, in which collision integrals involving unlike particles are
evaluated based on Maxwellian distributions, whereas collisions integrals involving like particles account for
non-Maxwellian effects.
\bigskip

\noindent \textit{Energy Distribution}
\bigskip

Similar to the assumption of a Maxwellian velocity distribution, the assumption of an equipartition of energy
(EE) has also been made periodically in an effort to simplify the evaluation of collision integrals. Again, an
equipartition of energy between unlike particles is only expected for a perfectly elastic system in equilibrium.
Numerous theoretical \cite{Garzo99, Barrat02, Montanero03}, simulation \cite{Clelland02, Dahl02,
MG02bis,Dahl2-02,Alam03, Paolotti03}, and experimental \cite{Feitosa02,Wildman02,SUKSS06} studies of granular
materials provide evidence that an equipartition of energy does not exist, and that the level of
non-equipartition of energy (nEE) increases as the restitution coefficient decreases (dissipation increases) and
as the mass ratio gets further from unity. The impact of such non-equipartition has recently been evaluated in
the context of species segregation. In particular, Galvin, Dahl and Hrenya \cite{Galvin05} have found that the
driving forces for species segregation that arise from non-equipartition are significant over a moderate range
of parameters for the case of a thermally-driven system. Furthermore, for the case of an intruder particle in
the presence of gravity, both Brey et al.\cite{Brey05}, Garz\'o \cite{G06}, and Yoon and Jenkins\cite{Yoon06}
have found that the direction of species segregation may reverse due to non-equipartition effects. Together,
these studies indicate the importance of non-equipartition in a variety of systems. The mixture theories of
JM87, H01, GD02, R03, IA05, and GHD include the effects of non-equipartition, while those of JM89, WA99, Z95,
and S06 do not account for non-equipartition.
\bigskip

\noindent\textit{Hydrodynamic Variables}
\bigskip

The appropriate choice of hydrodynamic variables to include in a hydrodynamic theory depends on the timescale
associated with a given variable \cite{Ferziger72}. First consider the case of a molecular gas, in which two
timescales are relevant -- the ``kinetic'' time scale and the ``hydrodynamic'' time scale. The kinetic time
scale is ``fast'' and representative of the time between collisions (i.e., mean free time). The hydrodynamic
time scale is ``slow'' and is set by the gradients in the system; this is the timescale over which macroscopic
variables change.

It is important to note that the velocity distribution function $f_i$ (and thus the Boltzmann and Enskog
equations) carries information with it on the kinetic time scale. However, the macroscopic variables ($n_i$,
$U$, and $T$ for the case of a molecular gas) correspond to velocity moments of $f_i$ that are
\textit{collisional invariants} -- i.e., $n_i$, $U$, and $T$ are conserved upon collision. As a result, they
will stay constant over a time on the order of a mean free path, and vary in time over the much longer
hydrodynamic time scale. It is this property which defines the appropriate (minimum set of) hydrodynamic
variables needed to describe a system.

To illustrate the above point, consider the time evolution of mixture
temperature ($T$) and species temperature ($T_i$), again for the case of a
molecular gas with more than one species present:
\begin{eqnarray*}
\frac{DT}{Dt} \!\! &=& \!\! \mbox{(energy flux)} \\
\frac{DT_i}{Dt} \!\! &=& \!\! \mbox{(energy flux)} + \left(
\begin{array}{l}
\mbox{collisional exchange (source/sink)} \\
\mbox{between unlike species}
\end{array}
\right)
\end{eqnarray*}

Note that $T$ is a collisional invariant (the total energy is conserved upon
collision and thus no ``source'' term is present), but that $T_i$ is not a
collisional invariant since the energy associated with a given species can
change upon collision with another species. Also note that the time scale of
the flux term is set by gradients (e.g., Fourier's law), whereas the time
scale of the source term is collisional in nature and thus relatively
``fast.'' Hence, the mixture temperature $T$ is characterized by one (slow)
timescale and the species temperature is characterized by two (slow and
fast) timescales. The corresponding physical picture is the following: each
species is expected to have a rapid relaxation after a few collisions to a
distribution near local equilibrium (local mixture temperature), and then
all temperatures evolve according to the slow hydrodynamic mode. Thus, the
relevant hydrodynamic (macroscopic) variable is $T$ since it is associated
with a conserved quantity; all other variables $(T_i)$ are enslaved to it.
Thus, solving a separate balance for $T_i$ is superfluous -- indeed this is
not done for molecular gases.

Next, consider a granular system in which the particles engage in
dissipative collisions. Three timescales need to be considered, as
illustrated by the balances below:
\begin{eqnarray*}
\frac{DT}{Dt}\!\! &=&\!\!\mbox{(energy flux)}+\left(
\begin{array}{l}
\mbox{inelastic} \\
\mbox{dissipation} \\
\mbox{(sink)}
\end{array}
\right)  \\
\frac{DT_{i}}{dt}\!\! &=&\!\!\mbox{(energy flux)}+\left(
\begin{array}{l}
\mbox{inelastic} \\
\mbox{dissipation} \\
\mbox{(sink)}
\end{array}
\right) +\left(
\begin{array}{l}
\mbox{collisional exchange (source/sink)} \\
\mbox{between unlike species}
\end{array}
\right)
\end{eqnarray*}
As with the molecular systems, the energy flux is characterized by a slow (hydrodynamic) time scale and the
collisional exchange between species is characterized by a fast (kinetic) time scale. In the granular system,
inelastic dissipation also occurs, and its corresponding time scale is not as obvious. More specifically, the
dissipation rate depends not only on the collisional frequency (kinetic time scale), but also on the value of
restitution coefficient. As $\alpha$ approaches unity, the dissipation rate goes to zero, and the time scale
becomes large (hydrodynamic time scale). In other words, it is unclear \textit{a priori} whether this term is
characterized by a fast or slow timescale.

The previous observation leads to the question: for moderate (smaller) values of $\alpha $, does the time scale
become fast enough to approach the kinetic time scale or is there still a separation of time scales? If the
latter is assumed (that the time scale associated with inelastic dissipation is much longer than that of the
kinetic time scale), Garz\'{o} and Dufty \cite {Garzo99} have shown that in a HCS two-component system, (i)~the
cooling rates associated with mixture and species temperatures are equal and (ii)~the ratio of species
temperature $(T_{1}/T_{2})$ remains constant -- i.e., all temperatures decay at the same rate but their ratios
remain constant. Further note that $T_{1}/T_{2}$ is not equal to one for unlike particles -- i.e., a
non-equipartition of energy is predicted. To test this assumption of the separation of timescales, MD
simulations of a binary mixture in HCS were performed by Dahl et al. \cite{Dahl2-02}. The simulation results
indicate that a constant value of $T_{1}/T_{2}$ is achieved after only a few collisions, and this behavior was
confirmed over a wide range of parameters.

As a result of the aforementioned finding that the timescale associated with inelastic dissipation is
hydrodynamic (slow) in nature, the physical picture is analogous to that of a molecular system Namely, each
species has a rapid relaxation to a local ``equilibrium'' state (now characterized by a constant value of
non-equipartition) -- HCS -- and is then enslaved to slow (hydrodynamic) evolution. Thus, the relevant
hydrodynamic variables are the same -- $n_i$, ${\bf U}$, and $T$ -- and the balance of additional variables
(such as $T_i$) would be superfluous.

The practical implications of using $n_{i}$, ${\bf U}$, and $T$ as the hydrodynamic variables in granular flows
are twofold. First, only a single balance for $T$ is needed, instead of separate balance for each $T_{i}$. The
reduction in the number of governing equations is expected to lead to a considerable decrease in computational
overhead. Second, the level of non-equipartition is in a state of local ``equilibrium'' due to the fast time
time scale of $T_{i}$, and thus $T_{1}/T_{2}$ depends on local values of flow field variables and particle
properties (mixture composition, $\alpha $, etc.). Hence, even though a non-equipartition of energy is indeed
present, it does not appear explicitly in the transport coefficients. Instead, its dependency on the flow field
variables is incorporated into the transport coefficients -- i.e., the effect of non-equipartition is
\textit{implicitly} contained in the transport coefficients. It can be solved for explicitly once the flow field
variables have been solved for using, for example, the relation derived by Garz\'{o} and Dufty \cite{Garzo99}.

Of the non-equipartition theories listed in Table \ref{CMH.Tab.2}, GD02 and GHD use $T$ as the hydrodynamic
variable for the granular energy field, whereas the other non-equipartition theories use $T_i$ (JM87, H01, R03,
IA05). The choice of hydrodynamic variable(s) is a non-issue for theories which invoke an equipartition
assumption (JM89, Z95, WA99, S06) since they inherently assume $T = T_i$ and thus use $T$ as the energy
variable. (It is worthwhile to note that S06 defines temperature differently than others, namely $T_S = 3T$.
Furthermore, IA06 uses a species temperature $T_{i, \mathit{IA}}$, which is defined in terms of velocity
fluctuations relative the mean species velocity, rather than the mass-averaged velocity used for $T$.)
\bigskip

\noindent\textit{Recapitulation}
\bigskip

Unlike previous Enskog-based (dense) theories, the current effort is based on an expansion about the homogeneous
cooling state and employs $n_i$, $U$, and $T$ as the hydrodynamic variables. The former extends the range of
validity to strong dissipation levels while the latter results in fewer balance equations, thereby reducing the
computational cost. This approach was first put forth by Garz\'{o} and Dufty \cite{Garzo02} who instead used the
Boltzmann equation as their starting kinetic equation; the current work thus extends the domain of applicability
from dilute to moderately dense flows.
\bigskip

\newcommand{\ubu}{\boldsymbol{\mathrm{U}}}
\begin{table}[htp]\def\baselinestretch{1}
\begin{center}
\caption{Hydrodynamic Descriptions of Granular Mixtures}\bigskip\bigskip \label{CMH.Tab.1}
\begin{tabular}{l|c|c|c|c|c|l}
\hline
                     &               &                 &                  &               &                & {\bf Pair}\\
                     &               &                 &                  &               &                & {\bf Correlation}\\
                     &               &                 & {\bf Number}     & {\bf Kinetic} & {\bf SET}      & {\bf Function}\\
{\bf References}      & {\bf Abbrev.} & {\bf Dimension} & {\bf of
Species} & {\bf Theory}  & {\bf vs.\ RET} & {\bf $\chi_{ij}^{(0)}$} \\
\hline \cite{Jenkins87}&JM87&2D and 3D&2&Enskog&SET&2D \cite{Jenkins87}, 3D \cite{Mansoori71}\\
\hline \cite{Jenkins89} and \cite{Arnarson98}&JM89&3D&2&Enskog&RET&3D \cite{Mansoori71}\\
\hline \cite{Zamankhan95}&Z95&3D&s&Enskog&RET& \cite{Carnahan69}\\
\hline \cite{Willits99} and \cite{Alam02}&WA99&2D&2&Enskog&RET&2D \cite{Jenkins87}\\
\hline \cite{Huilin01}&H01&3D&2&Enskog&RET&\cite{Mansoori71}\\
\hline \cite{Garzo02}&GD02&3D&2&Boltzmann&--&--\\
\hline \cite{Rahaman03}&R03&3D&2&Enskog&SET&\cite{Mathiesen00}\\
\hline \cite{Iddir05}&IA05&3D&s&Enskog&SET&\cite{Lebowitz64}\\
\hline \cite{Serero06}&S06&3D&2&Boltzmann&--&--\\
\hline current work &GHD&2D and 3D&s&Enskog&RET&2D \cite{Luding04}, 3D \cite{Boublik70}
\end{tabular}
\end{center}
\end{table}

\begin{table}[htp]\def\baselinestretch{1}
\caption{Hydrodynamic Descriptions of Granular Mixtures}\bigskip\bigskip \label{CMH.Tab.2}
\begin {tabular}{c|c|c|c|c|c|c|c}\hline
           &                & {\bf Chapman-}  & {\bf Chapman-}   &                 & {\bf Single}      &                   & \\
           &                & {\bf Enskog}    & {\bf Enskog}     & {\bf Sonine}    & {\bf Particle}    &                   & {\bf Hydro-}    \\
           & {\bf Solution} & {\bf Expansion} & {\bf Expansion}  & {\bf Expansion} & {\bf Velocity}    & {\bf Energy}      & {\bf dynamic}   \\
{\bf Ref.} & {\bf Method}   & {\bf Order}     & {\bf Base State} & {\bf Order}     & {\bf Distribution} & {\bf
Distributon} & {\bf
Variables} \\
\hline
JM87& CE& NS & $\alpha=1$ & 1st &M&nEE& $n_i$, $\ubu_i$, $T_i$\\
\hline
JM89& CE& NS & $\alpha=1$ & 1st&nM&EE& $n_i$, $\ubu_i$, $T$\\
\hline
Z95&Grad's method&--& --& --&nM&EE& $13$ moments\\
\hline
WA99& CE& NS & $\alpha=1$ & 1st&nM &EE& $n_i$, $\ubu_i$, $T$\\
\hline
H01& CE& NS & $\alpha=1$ & 1st&M&nEE& $n_i$, $\ubu_i$, $T_i$\\
\hline
GD02& CE& NS & HCS& 1st&nM&nEE& $n_i$, $\ubu_i$, $T$\\
\hline
R03& CE& NS & $\alpha=1$ & 1st&M&nEE& $n_i$, $\ubu_i$, $T_{i}$\\
\hline
IA05& CE& NS &$\alpha=1$& 1st&M/nM&nEE& $n_i$, $\ubu_i$, $T_{i,IA}$\\
\hline
S06& CE& NS & $\alpha=1$ &3rd&nM&EE& $n_i$, $\ubu_i$, $T_s$\\
\hline GHD& CE& NS & HCS& 1st&nM&nEE& $n_i$, $\ubu_i$, $T$
\end{tabular}
\end{table}

\acknowledgments

V. G. acknowledges partial support from the Ministerio de Ciencia y Tecnolog\'{\i}a (Spain) through Grant No.
FIS2007--60977. C.M.H. is grateful to the National Science Foundation for providing financial support of this
project through grant CTS-0318999 with additional support provided by the American Chemical Society Petroleum
Research Fund (Grant 43393-AC9) and the Engineering and Physical Sciences Research Council (Grant EP/DO30676/1).
C. M. H. and J. W. D. are also grateful to the organizers and participants of the Granular Physics Workshop at
the Kavli Institute of Theoretical Physics (with partial support from the National Science Foundation under
grant PHY99-07949), which provided a starting forum for much of this work.

\appendix
\section{Collision frequencies}
\label{appA}

In this Appendix we display the expressions for the collision frequencies appearing in the evaluation of the
kinetic contributions to the transport coefficients and in the first order contributions to the cooling rate. As
noted in the main text, most of these frequencies (those corresponding to the transport coefficients) have been
obtained in the low-density limit (except for the factors $\chi _{ij}^{(0)}$) \cite{GM06} for arbitrary
dimensions by considering the Maxwellian approximation for the zeroth-order distribution functions
$f_{i}^{(0)}$. Their expressions are
\begin{equation}
\nu _{ii}=\frac{2\pi ^{(d-1)/2}}{d\Gamma \left( \frac{d}{2}\right) } \sum_{j\neq i}^{s}n_{j}\sigma
_{ij}^{d-1}\chi _{ij}^{(0)}\mu _{ji}v_{0}(1+\alpha _{ij})\left( \frac{\theta _{i}+\theta _{j}}{\theta _{i}\theta
_{j}}\right) ^{1/2},  \label{a1}
\end{equation}
\begin{equation}
\nu _{ij}=-\frac{2\pi ^{(d-1)/2}}{d\Gamma \left( \frac{d}{2}\right) }n_{i} \sigma _{ij}^{d-1}\chi _{ij}^{(0)}\mu
_{ij}v_{0}(1+\alpha _{ij})\left( \frac{\theta _{i}+\theta _{j}}{\theta _{i}\theta _{j}}\right) ^{1/2},
\label{a1.1}
\end{equation}
\begin{eqnarray}
\tau _{ii} &=&\frac{2\pi ^{(d-1)/2}}{d(d+2)\Gamma \left( \frac{d}{2}\right) } v_{0}\left\{ n_{i}\sigma
_{i}^{d-1}\chi _{ii}^{(0)}(2\theta _{i})^{-1/2}(3+2d-3\alpha _{ii})(1+\alpha _{ii})\right.  \notag  \label{a2}
\\
&&+2\sum_{j\neq i}^{s}n_{j}\chi _{ij}^{(0)}\sigma _{ij}^{d-1}\mu
_{ji}(1+\alpha _{ij})\theta _{i}^{3/2}\theta _{j}^{-1/2}\left[ (d+3)\beta
_{ij}\theta _{i}^{-2}(\theta _{i}+\theta _{j})^{-1/2}\right.  \notag \\
&&\left. \left. +\frac{3+2d-3\alpha _{ij}}{2}\mu _{ji}\theta
_{i}^{-2}(\theta _{i}+\theta _{j})^{1/2}+\frac{2d(d+1)-4}{2(d-1)}\theta
_{i}^{-1}(\theta _{i}+\theta _{j})^{-1/2}\right] \right\} ,
\end{eqnarray}
\begin{eqnarray}
\tau _{ij} &=&\frac{4\pi ^{(d-1)/2}}{d(d+2)\Gamma \left( \frac{d}{2}\right) } v_{0}n_{i}\chi _{ij}^{(0)}\sigma
_{ij}^{d-1}\mu _{ij}\theta
_{j}^{3/2}\theta _{i}^{-1/2}(1+\alpha _{ij})  \notag  \label{a3} \\
&\times &\left[ (d+3)\beta _{ij}\theta _{j}^{-2}(\theta _{i}+\theta
_{j})^{-1/2}+\frac{3+2d-3\alpha _{ij}}{2}\mu _{ji}\theta _{j}^{-2}(\theta
_{i}+\theta _{j})^{1/2}\right.  \notag \\
&&\left. -\frac{2d(d+1)-4}{2(d-1)}\theta _{j}^{-1}(\theta _{i}+\theta
_{j})^{-1/2}\right] ,
\end{eqnarray}
\begin{eqnarray}
\omega _{ii} &=&\frac{\pi ^{(d-1)/2}}{\Gamma \left( \frac{d}{2}\right) } \frac{2}{d\sqrt{2}}\sigma
_{i}^{d-1}n_{i}\chi _{ii}^{(0)}v_{0}\theta
_{i}^{-1/2}(1-\alpha _{ii}^{2})  \notag  \label{a4} \\
&&+\frac{\pi ^{(d-1)/2}}{\Gamma \left( \frac{d}{2}\right) }\frac{2}{d(d+2)} \sum_{j\neq i}^{s}n_{j}\chi
_{ij}^{(0)}\sigma _{ij}^{d-1}v_{0}\mu _{ji}(1+\alpha _{ij})(\theta _{i}+\theta _{j})^{-1/2}\theta
_{i}^{1/2}\theta _{j}^{-3/2}A_{ij},
\end{eqnarray}
\begin{equation}
\omega _{ij}=\frac{\pi ^{(d-1)/2}}{\Gamma \left( \frac{d}{2}\right) }\frac{2 }{d(d+2)}n_{j}\chi
_{ij}^{(0)}\sigma _{ij}^{d-1}v_{0}\mu _{ji}(1+\alpha _{ij})(\theta _{i}+\theta _{j})^{-1/2}\theta
_{i}^{1/2}\theta _{j}^{-3/2}C_{ij},  \label{a5}
\end{equation}
\begin{eqnarray}
\gamma _{ii} &=&\frac{\pi ^{(d-1)/2}}{\Gamma \left( \frac{d}{2}\right) } \frac{8}{d(d+2)}\sigma
_{i}^{d-1}n_{i}\chi _{ii}^{(0)}v_{0}(2\theta _{i})^{-1/2}(1+\alpha _{ii})\left[
\frac{d-1}{2}+\frac{3}{16}(d+8)(1-\alpha
_{ii})\right]  \notag  \label{a6} \\
&&+\frac{\pi ^{(d-1)/2}}{\Gamma \left( \frac{d}{2}\right) }\frac{1}{d(d+2)} \sum_{j\neq i}^{s}n_{j}\chi
_{ij}^{(0)}\sigma _{ij}^{d-1}v_{0}\mu _{ji}(1+\alpha _{ij})\left( \frac{\theta _{i}}{\theta _{j}(\theta
_{i}+\theta _{j})}\right) ^{3/2}\left[ E_{ij}-(d+2)\frac{\theta _{i}+\theta
_{j}}{\theta _{i}}A_{ij}\right] ,  \notag \\
&&
\end{eqnarray}
\begin{equation}
\gamma _{ij}=-\frac{\pi ^{(d-1)/2}}{\Gamma \left( \frac{d}{2}\right) }\frac{1 }{d(d+2)}n_{i}\chi
_{ij}^{(0)}\sigma _{ij}^{d-1}v_{0}\mu _{ij}(1+\alpha _{ij})\left( \frac{\theta _{j}}{\theta _{i}(\theta
_{i}+\theta _{j})}\right) ^{3/2}\left[ F_{ij}+(d+2)\frac{\theta _{i}+\theta _{j}}{\theta _{j}}C_{ij} \right] .
\label{a7}
\end{equation}
In the above equations, $v_{0}=\sqrt{2T/m}$ is the thermal velocity, $\theta
_{i}=m_{i}T/mT_{i}$, and we have introduced the dimensionless quantities
\begin{eqnarray}
A_{ij} &=&(d+2)(2\beta _{ij}+\theta _{j})+\mu _{ji}(\theta _{i}+\theta
_{j})\left\{ (d+2)(1-\alpha _{ij})-[(11+d)\alpha _{ij}-5d-7]\beta
_{ij}\theta _{i}^{-1}\right\}  \notag  \label{a8} \\
&&+3(d+3)\beta _{ij}^{2}\theta _{i}^{-1}+2\mu _{ji}^{2}\left( 2\alpha
_{ij}^{2}-\frac{d+3}{2}\alpha _{ij}+d+1\right) \theta _{i}^{-1}(\theta
_{i}+\theta _{j})^{2}-(d+2)\theta _{j}\theta _{i}^{-1}(\theta _{i}+\theta
_{j}),  \notag \\
&&
\end{eqnarray}
\begin{eqnarray}
C_{ij} &=&(d+2)(2\beta _{ij}-\theta _{i})+\mu _{ji}(\theta _{i}+\theta
_{j})\left\{ (d+2)(1-\alpha _{ij})+[(11+d)\alpha _{12}-5d-7]\beta
_{ij}\theta _{j}^{-1}\right\}  \notag  \label{a9} \\
&&-3(d+3)\beta _{ij}^{2}\theta _{j}^{-1}-2\mu _{ji}^{2}\left( 2\alpha
_{ij}^{2}-\frac{d+3}{2}\alpha _{ij}+d+1\right) \theta _{j}^{-1}(\theta
_{i}+\theta _{j})^{2}+(d+2)(\theta _{i}+\theta _{j}),  \notag \\
&&
\end{eqnarray}
\begin{eqnarray}
E_{ij} &=&2\mu _{ji}^{2}\theta _{i}^{-2}(\theta _{i}+\theta _{j})^{2}\left(
2\alpha _{ij}^{2}-\frac{d+3}{2}\alpha _{ij}+d+1\right) \left[ (d+2)\theta
_{i}+(d+5)\theta _{j}\right]  \notag  \label{a10} \\
&&-\mu _{ji}(\theta _{i}+\theta _{j})\left\{ \beta _{ij}\theta
_{i}^{-2}[(d+2)\theta _{i}+(d+5)\theta _{j}][(11+d)\alpha _{ij}-5d-7]\right.
\notag \\
&&\left. -\theta _{j}\theta _{i}^{-1}[20+d(15-7\alpha _{ij})+d^{2}(1-\alpha
_{ij})-28\alpha _{ij}]-(d+2)^{2}(1-\alpha _{ij})\right\}  \notag \\
&&+3(d+3)\beta _{ij}^{2}\theta _{i}^{-2}[(d+2)\theta _{i}+(d+5)\theta
_{j}]+2\beta _{ij}\theta _{i}^{-1}[(d+2)^{2}\theta _{i}+(24+11d+d^{2})\theta
_{j}]  \notag \\
&&+(d+2)\theta _{j}\theta _{i}^{-1}[(d+8)\theta _{i}+(d+3)\theta
_{j}]-(d+2)(\theta _{i}+\theta _{j})\theta _{i}^{-2}\theta _{j}[(d+2)\theta
_{i}+(d+3)\theta _{j}],  \notag \\
&&
\end{eqnarray}
\begin{eqnarray}
F_{ij} &=&2\mu _{ji}^{2}\theta _{j}^{-2}(\theta _{i}+\theta _{j})^{2}\left(
2\alpha _{ij}^{2}-\frac{d+3}{2}\alpha _{ij}+d+1\right) \left[ (d+5)\theta
_{i}+(d+2)\theta _{j}\right]  \notag  \label{a11} \\
&&-\mu _{ji}(\theta _{i}+\theta _{j})\left\{ \beta _{ij}\theta
_{j}^{-2}[(d+5)\theta _{i}+(d+2)\theta _{j}][(11+d)\alpha _{ij}-5d-7]\right.
\notag \\
&&\left. +\theta _{i}\theta _{j}^{-1}[20+d(15-7\alpha _{ij})+d^{2}(1-\alpha
_{ij})-28\alpha _{ij}]+(d+2)^{2}(1-\alpha _{ij})\right\}  \notag \\
&&+3(d+3)\beta _{ij}^{2}\theta _{j}^{-2}[(d+5)\theta _{i}+(d+2)\theta
_{j}]-2\beta _{ij}\theta _{j}^{-1}[(24+11d+d^{2})\theta _{i}+(d+2)^{2}\theta
_{j}]  \notag \\
&&+(d+2)\theta _{i}\theta _{j}^{-1}[(d+3)\theta _{i}+(d+8)\theta
_{j}]-(d+2)(\theta _{i}+\theta _{j})\theta _{j}^{-1}[(d+3)\theta
_{i}+(d+2)\theta _{j}],  \notag \\
&&
\end{eqnarray}
where $\beta _{ij}\equiv \mu _{ij}\theta _{j}-\mu _{ji}\theta _{i}$.

It only remains to evaluate the collision frequencies $ \psi_{ij} $ needed to get the first order contribution
$\zeta^{(1,1)}$ to the cooling rate. These frequencies have not been previously determined in the Boltzmann
limit. To compute them, we use the property
\begin{eqnarray}
\int d\mathbf{v}_{1}h(\mathbf{v}_{1})J_{ij}^{(0}\left[ \mathbf{V} _{1}|\Phi_{i},\Phi_{j}\right] &=&
\chi_{ij}^{(0)}\sigma_{ij}^{d-1} \int \,d \mathbf{v}_{1}\,\int \,d\mathbf{v}_{2}\Phi_{i}(
\mathbf{v}_{1})\Phi_{j}(
\mathbf{v}_{2})  \notag \\
&&\times \int d\widehat{\boldsymbol{\sigma }}\,\Theta (\widehat{\boldsymbol{ \sigma}} \cdot
\mathbf{g})(\widehat{\boldsymbol{\sigma }}\cdot \mathbf{g})\, \left[ h(
\mathbf{V}_{1}^{^{\prime}})-h(\mathbf{V}_{1})\right] \;, \label{a12}
\end{eqnarray}
with
\begin{equation}
\mathbf{V}_{1}^{\prime }=\mathbf{V}_{1}-\mu _{ji}(1+\alpha _{ij})( \widehat{ \boldsymbol{\sigma }}\cdot
\mathbf{g})\widehat{\boldsymbol{\sigma}}\;. \label{a13}
\end{equation}
To determine $\psi_{ij}$, let us consider first the integral
\begin{equation}  \label{a14}
I_\psi=\int\, d\mathbf{v}\; \left(\frac{m_iV^2}{2T_i}\right)^2\;
J_{ij}^{(0)}[f_i^{(0)},f_{j,M}F_j].
\end{equation}
Use of Eqs.\ (\ref{a12}) and (\ref{a13}) gives
\begin{equation}  \label{a15}
I_\psi=\chi_{ij}^{(0)}\sigma_{ij}^{d-1}\frac{m_i^2}{4T_i^2} \int \,d\mathbf{v }_{1}\,\int
\,d\mathbf{v}_{2}f_{i}^{(0)}( \mathbf{V}_{1})f_{j,M}(\mathbf{V} _{2})F_j(\mathbf{V}_2) \int
d\widehat{\boldsymbol{\sigma }}\,\Theta ( \widehat{\boldsymbol{\sigma}} \cdot
\mathbf{g})(\widehat{\boldsymbol{\sigma } }\cdot \mathbf{g})\, \left(V_1^{\prime 4}-V_1^4 \right) \;.
\end{equation}
The collision rule (\ref{a13}) yields
\begin{eqnarray}  \label{b18}
V_1^{\prime 4}-V_1^4&=&2\mu_{ji}^2(1+\alpha_{ij})^2(\widehat{\boldsymbol{ \sigma }}\cdot \mathbf{g})^2
\left[2(\widehat{\boldsymbol{\sigma }}\cdot \mathbf{V}_{1})^2+V_1^2+\frac{\mu_{ji}^2}{2}
(1+\alpha_{ij})^2(\widehat{
\boldsymbol{\sigma }}\cdot \mathbf{g})^2\right]  \notag \\
& & -4\mu_{ji}(1+\alpha_{ij})(\widehat{\boldsymbol{\sigma }}\cdot \mathbf{g} ) (\widehat{\boldsymbol{\sigma
}}\cdot \mathbf{V}_{1})\left[V_1^2+ \mu_{ji}^2(1+\alpha_{ij})^2(\widehat{\boldsymbol{\sigma }}\cdot
\mathbf{g})^2 \right].
\end{eqnarray}
The integration over the solid angle in (\ref{a15}) leads to
\begin{eqnarray}  \label{b7}
\int d\widehat{\boldsymbol{\sigma }}\,\Theta (\widehat{\boldsymbol{\sigma}} \cdot
\mathbf{g})(\widehat{\boldsymbol{\sigma }}\cdot \mathbf{g})\, \left(V_1^{\prime
4}-V_1^4\right)&=&\frac{4B_3}{d+3} \mu_{ji}^2(1+ \alpha_{ij})^2g\left[3(\mathbf{V}_1\cdot
\mathbf{g})^2+\frac{d+5}{2}
g^2V_1^2+ \mu_{ji}^2(1+\alpha_{ij})^2g^4\right]  \notag \\
& & -\frac{4B_3}{d+3} \mu_{ji}(1+\alpha_{ij})g(\mathbf{V}_1\cdot \mathbf{g})
\left[(d+3)V_1^2+4 \mu_{ji}^2(1+\alpha_{ij})^2g^2\right],  \notag \\
\end{eqnarray}
where
\begin{equation}  \label{b6}
B_k\equiv \int d\widehat{\boldsymbol{\sigma}}\, \Theta (\widehat{{ \boldsymbol{\sigma }}} \cdot \mathbf{g})\,
(\widehat{\boldsymbol{\sigma}} \cdot {\widehat{\mathbf{g}}})^k=\pi^{(d-1)/2} \frac{ \Gamma\left(\frac{k+1}{2
}\right)}{\Gamma\left(\frac{k+d}{2}\right)}.
\end{equation}
Therefore, the integral (\ref{a15}) can be written as
\begin{equation}  \label{a16}
I_\psi=\frac{B_3}{d+3}\chi_{ij}^{(0)}\sigma_{ij}^{d-1}n_in_j(\theta_i \theta_j)^{d/2}v_0
\mu_{ji}(1+\alpha_{ij})\theta_i^2\Delta_j I_\psi^*(\theta_i,\theta_j),
\end{equation}
where we have taken the Maxwellian form for $f_i^{(0)}$ and have introduced
the dimensionless integral
\begin{eqnarray}  \label{a17}
I_{\psi}^*(\theta_i,\theta_j)&=&\pi^{-d} \int \,d\mathbf{c}_{1}\,\int \,d \mathbf{c}_{2}\,\, e^{-(\theta
_{i}c_{1}^{2}+\theta_{j}c_{2}^{2})}x  \notag
\\
& &\times \left\{ 4 \mu_{ji}(1+\alpha_{ij})\left[3(\mathbf{c}_1\cdot \mathbf{ x})^2+\frac{d+5}{2}g^2c_1^2+
\mu_{ji}^2(1+\alpha_{ij})^2x^4\right]\right.
\notag \\
& & \left.-4 (\mathbf{c}_1\cdot \mathbf{x})\left[(d+3)c_1^2+4
\mu_{ji}^2(1+\alpha_{ij})^2x^2\right]\right\},
\end{eqnarray}
and the operator
\begin{equation}  \label{a18}
\Delta_j\equiv \theta_j^2\frac{\partial^2}{\partial \theta_j^2}+(d+2)\theta_j \frac{\partial}{\partial
\theta_j}+\frac{d(d+2)}{4}.
\end{equation}
In addition, $\mathbf{c}_i\equiv \mathbf{V}_1/v_0$ and $\mathbf{x}\equiv
\mathbf{g}/v_0$. The integral (\ref{a17}) can be performed by the change of
variables $\{\mathbf{c}_1, \mathbf{c}_2\}\to \{\mathbf{x}, \mathbf{y}\}$
where $\mathbf{y}\equiv \theta_i\mathbf{c}_1+\theta_j\mathbf{c}_2$ and the
Jacobian is $(\theta_i+\theta_j)^{-d}$. With this change the integrations
can be done quite efficiently by using a computer package of symbolic
calculations. A lengthy and careful algebra gives
\begin{eqnarray}  \label{a19}
I_{\psi}^*(\theta_i,\theta_j)&=&(d+3) \frac{\Gamma\left(\frac{d+3}{2}\right) }
{\Gamma\left(\frac{d}{2}\right)}\left(\theta_i\theta_j\right)^{-d/2}
(\theta_i\theta_j )^{-5/2} (\theta_i+\theta_j)^{-1/2}  \notag \\
& & \times \left\{ -2\theta_j^2\left[(d+3)\theta_j+(d+2)\theta_{i}\right]
+\mu _{ji}\left( 1+\alpha _{ij}\right) \theta_j\left(\theta_i+\theta_j
\right)\right.  \notag \\
& &\times \left[ (11+ d)\theta_j+\frac{d^2+5d+6}{d+3} \theta_{i} \right]
\notag \\
& & \left. -8\mu _{ji}^{2}\left( 1+\alpha _{ij}\right)
^{2}\theta_j\left(\theta_i+\theta_j \right) ^{2} +2\mu _{ji}^{3}\left(
1+\alpha _{ij}\right) ^{3}\left( \theta_i+\theta_j \right) ^{3}\right\} \;.
\end{eqnarray}
Use of (\ref{a16}) in (\ref{a19}) leads to the final expression for $I_\psi$ :
\begin{eqnarray}
\label{a20}
I_{\psi} &=&\frac{\pi^{(d-1)/2}} {4\Gamma\left(\frac{d}{2}\right)}
n_in_j\chi_{ij}^{(0)}\sigma_{ij}^{d-1}v_0\mu _{ji}\left( 1+\alpha
_{ij}\right)\theta_i^{3/2}\theta_j^{-3/2}\left(\theta_i+\theta_j \right)
^{-5/2}  \notag \\
& & \times\left\{ \left[ (d-1)\theta_j+(d+2)\theta_{i} \right]\left[ 2\theta_j +3\mu _{ji}\left( 1+\alpha
_{ij}\right) \left(\theta_i+\theta_j
\right)\right] \right.  \notag \\
& & \left. -24\mu _{ji}^{2}\left( 1+\alpha _{ij}\right)
^{2}\left(\theta_i+\theta_j \right)^2+30\mu _{ji}^{3}\left( 1+\alpha
_{ij}\right) ^{3}\theta_j^{-1}\left(\theta_i+\theta_j \right)^3\right\}.
\end{eqnarray}

Following similar mathematical steps as made before for $ I_{\psi }$, one obtains
\begin{eqnarray}
\int d\mathbf{v}\frac{m_{i}V^{2}}{2T_{i}} J_{ij}^{(0)}[f_{i}^{(0)},f_{j,M}F_{j}] &=&\frac{\pi
^{(d-1)/2}}{4\Gamma \left( \frac{d}{2}\right) }n_{i}n_{j}\chi _{ij}^{(0)}\sigma _{ij}^{d-1}v_{0}\mu _{ji}\left(
1+\alpha _{ij}\right) \theta _{i}^{3/2}\theta _{j}^{-3/2}\left( \theta _{i}+\theta _{j}\right) ^{-3/2}
\notag  \label{a21} \\
&&\times \left[ 2\theta _{j}+3\mu _{ji}\left( 1+\alpha _{ij}\right) (\theta
_{i}+\theta _{j})\right] .
\end{eqnarray}
Combining Eqs.\ (\ref{a20}) and (\ref{a21}), one gets the result
\begin{eqnarray}
\int d\mathbf{v}F_{i}(V)J_{ij}^{(0)}[f_{i}^{(0)},f_{j,M}F_{j}] &=&\frac{\pi
^{(d-1)/2}}{4\Gamma \left( \frac{d}{2}\right) }n_{i}n_{j}\chi
_{ij}^{(0)}\sigma _{ij}^{d-1}v_{0}\mu _{ji}\left( 1+\alpha _{ij}\right)
\theta _{i}^{3/2}\theta _{j}^{-3/2}(\theta _{i}+\theta _{j})^{-5/2}  \notag
\label{a22} \\
&&\times \left\{ \left[ (d-1)\theta _{j}+(d+2)\theta _{i}\right] \left[
2\theta _{j}+3\mu _{ji}\left( 1+\alpha _{ij}\right) \left( \theta
_{i}+\theta _{j}\right) \right] \right.   \notag \\
&&-24\mu _{ji}^{2}\left( 1+\alpha _{ij}\right) ^{2}\left( \theta _{i}+\theta
_{j}\right) ^{2}+30\mu _{ji}^{3}\left( 1+\alpha _{ij}\right) ^{3}\theta
_{j}^{-1}\left( \theta _{i}+\theta _{j}\right) ^{3}  \notag \\
&&\left. -(d+2)\left( \theta _{i}+\theta _{j}\right) \left[ 2\theta
_{j}+3\mu _{ji}\left( 1+\alpha _{ij}\right) (\theta _{i}+\theta _{j})\right]
\right\} .
\end{eqnarray}
The remaining integrals needed to determine the collision frequencies $\psi
_{ii}$ and $\psi _{ij}$ can be also obtained by performing identical
mathematical steps. Their expressions are
\begin{eqnarray}
\label{a23} \int d{\bf v}F_i({\bf V})J_{ij}^{(0)}[f_{i,M}F_i,f_j^{(0)}]&=&\frac{\pi^{(d-1)/2}}
{4\Gamma\left(\frac{d}{2}\right)} n_in_j\chi_{ij}^{(0)}\sigma_{ij}^{d-1}v_0 \mu_{ji}\left( 1+\alpha
_{ij}\right)(\theta_i\theta_j)^{-1/2}(\theta_i+\theta_j)^{-5/2}
\nonumber\\
& &\times\left\{-2\left[
(45+15d)\theta_j^3+3(38+13d)\theta_{i}\theta_j^2\right.\right.\nonumber\\
& & \left.
+8(11+4d)\theta_i^2\theta_{j}+8(2+d)\theta_{i} ^{3}\right] \nonumber \\
&&+3\mu _{ji}\left( 1+\alpha _{ij}\right) \left(\theta_i+\theta_{j} \right) \left[
(55+5d)\theta_j^2+9(10+d)\theta_i\theta_{j} +4(8+d)\theta_{i}^2
\right]\nonumber\\
& &  -24\mu _{ji}^{2}\left( 1+\alpha _{ij}\right) ^{2}\left( \theta_i+\theta_{j} \right) ^{2}\left(
5\theta_j+4\theta_{i} \right)+30\mu _{ji}^{3}\left( 1+\alpha _{ij}\right) ^{3}\left(
\theta_i+\theta_{j} \right) ^{3}\nonumber\\
& & \left.+(d+2)\theta_j\left(\theta_i+\theta_j \right) \left[2(4\theta_i+3\theta_j)-3\mu_{ji}\left( 1+\alpha
_{ij}\right)(\theta_i+\theta_j)\right]\right\},\nonumber\\
\end{eqnarray}
\begin{eqnarray}
\label{a24} \int d{\bf v}F_i({\bf
V})J_{ii}^{(0)}[f_i^{(0)},f_{i,M}F_i]&=&\frac{3\sqrt{2}}{64}\frac{\pi^{(d-1)/2}}
{\Gamma\left(\frac{d}{2}\right)}n_i^2\chi_{ii}^{(0)}\sigma_{i}^{d-1} \left( 1+\alpha
_{ij}\right)v_0\theta_i^{-1/2}\nonumber\\
& & \times\left[10\alpha_{ii}^3+22\alpha_{ii}^2+11\alpha_{ii}-3\right],
\end{eqnarray}
\begin{eqnarray}
\label{a24.1} \int d{\bf v}F_i({\bf
V})J_{ii}^{(0)}[f_{i,M}F_i,f_i^{(0)}]&=&\frac{\sqrt{2}}{64}\frac{\pi^{(d-1)/2}}
{\Gamma\left(\frac{d}{2}\right)}n_i^2\chi_{ii}^{(0)}\sigma_{i}^{d-1} \left( 1+\alpha
_{ij}\right)v_0\theta_i^{-1/2}\nonumber\\
& & \times\left[30\alpha_{ii}^3-126\alpha_{ii}^2+177\alpha_{ii}+16d(3\alpha_{ii}-7)-137\right].
\nonumber\\
\end{eqnarray}
The expressions for the frequencies $\psi _{ii}$ and $\psi _{ij}$ can be easily obtained from Eqs.\
(\ref{a23})--(\ref{a25}) when one takes into account their definitions (\ref{1.14}) and (\ref{1.15}).

In the case of mechanically equivalent particles ($ \chi_{ij}^{(0)}=\chi^{(0)}$, $\sigma_{i}=\sigma$, and
$\alpha_{ij}=\alpha$), Eqs. (\ref{a24}) and (\ref{a24.1}) yield
\begin{eqnarray}  \label{a25}
\int d\mathbf{v}F(\mathbf{V})\left(J^{(0)}[f^{(0)},f_{M}F(\mathbf{V}
)]+J^{(0)}[f_{M}F(\mathbf{V}),f^{(0)}]\right)&=&\frac{\pi^{(d-1)/2}} {
8\Gamma\left(\frac{d}{2}\right)}\chi^{(0)}\sigma^{d-1}n_in_j\frac{1+\alpha}{2
}  \notag \\
& \times& \left( 30\alpha^3-30\alpha^2+105\alpha+24d\alpha-56d-73\right).
\notag \\
\end{eqnarray}
This expression coincides with the one previously derived for a monodisperse granular gas \cite{Lutsko05}.

\section{Collision integrals}
\label{appB}

In this Appendix, we provide some of the mathematical steps to compute the different collision integrals
involving the operator ${\cal K}_{ij,\gamma}[X]$. This operator is defined as
\begin{eqnarray}
\mathcal{K}_{ij,\gamma }[X_{j}] &=&\sigma _{ij}^{d}\chi _{ij}^{(0)}\int d \mathbf{v}_{2}\int
d\widehat{\boldsymbol {\sigma }}\Theta (\widehat{\boldsymbol {\sigma}} \cdot
\mathbf{g}_{12})(\widehat{\boldsymbol {\sigma }}\cdot \mathbf{g}_{12})
\widehat{{\sigma}}_\gamma  \nonumber\\
&&\times \left[ \alpha _{ij}^{-2}f_{i}^{(0)}(\mathbf{V} _{1}^{\prime \prime})X_{j}(\mathbf{V}_{2}^{\prime
\prime} )+f_{i}^{(0)}(\mathbf{V}_{1})X_{j}(\mathbf{V}_{2})\right] . \label{b0}
\end{eqnarray}
To simplify all these type of integrals, we use the property
\begin{eqnarray}
\int d{\bf v}_{1}h({\bf V}_{1}){\cal K}_{ij,\gamma}[X_j({\bf V}_{2})] &=&-\chi_{ij}^{(0)}\sigma _{ij}^{d}\int
d{\bf v}_{1}\int d{\bf
v}_{2}f_{i}^{(0)}( {\bf V}_{1})X_{j}({\bf V}_{2}) \nonumber\\
& & \times\int d\widehat{\boldsymbol {\sigma }}\,\Theta (\widehat{\boldsymbol {\sigma}} \cdot {\bf
g})(\widehat{\boldsymbol {\sigma }}\cdot {\bf g})\widehat{\sigma}_\gamma \left[ h( {\bf V}_{1}')-h({\bf
V}_{1})\right] \;,  \label{b1}
\end{eqnarray}
where ${\bf V}_1'$ is defined by Eq.\ (\ref{a13}).

Let us start with the collision integrals appearing in the evaluation of the mass flux. One of them is
\begin{equation}
\label{b3} I_D\equiv \int\; d{\bf v}_1 m_iV_{1,\gamma}{\cal K}_{ij,\gamma}[\nabla_{{\bf V}_2}\cdot ({\bf
V}_2f_j^{(0)})].
\end{equation}
Use of the identity (\ref{b1}) in (\ref{b3}) gives
\begin{equation}
I_D= B_2\sigma _{ij}^{d}\chi_{ij}^{(0)} m_i\mu_{ji}(1+\alpha _{ij}) \int \,d{\bf V}_{1}\,\int \,d{\bf V}_{2}
f_{i}^{(0)}({\bf V}_{1})\nabla_{{\bf V}_2}\cdot \left({\bf V}_2f_j^{(0)}({\bf V}_2)\right)g^2.  \label{b4}
\end{equation}
The integral (\ref{b4}) can be exactly evaluated and the result is
\begin{equation}
\label{b7.1} I_D=-2d B_2 n_in_j \sigma _{ij}^{d}\chi_{ij}^{(0)}\mu_{ij}(1+\alpha _{ij})T_j.
\end{equation}
The remaining integrals corresponding to the mass flux can be computed by using similar mathematical steps as
those made before for $I_D$. The results are
\begin{equation}
\label{b8} \int\; d{\bf v}_1 m_i V_{1,\beta}{\cal K}_{ij,\beta}[f_j^{(0)}]=d B_2 n_in_j \sigma
_{ij}^{d}\chi_{ij}^{(0)}m_i \mu_{ji}(1+\alpha _{ij})\left(\frac{T_i}{m_i}+\frac{T_j}{m_j}\right),
\end{equation}
\begin{eqnarray}
\label{b9} \int\; d{\bf v}_1 m_i V_{1,\beta}{\cal K}_{i\ell,\beta}[n_j\partial_{n_j}f_\ell^{(0)}]&=& \int\;
d{\bf v}_1 m_iV_{1,\beta}\left({\cal K}_{i\ell,\beta}[\delta_{j\ell} f_\ell^{(0)}]-\frac{1}{2}{\cal
K}_{i\ell,\beta}[\nabla_{{\bf V}_2}\cdot ({\bf V}_2f_\ell^{(0)})]n_j\partial_{n_j}\ln
\gamma_\ell \right)\nonumber\\
&=&dB_2n_in_j \sigma _{i\ell}^{d}\chi_{i\ell}^{(0)}m_i \mu_{\ell i}(1+\alpha
_{i\ell})\left[\delta_{j\ell}\left(\frac{T_i}{m_i}+\frac{T_\ell}{m_\ell}\right)\right.
\nonumber\\
& & \left. +\frac{n_\ell T_\ell}{n_jm_\ell}\frac{\partial \ln \gamma_\ell}{\partial \ln n_j}\right].
\end{eqnarray}

The collision integral involved in the evaluation of the pressure tensor is of the form
\begin{eqnarray}
\label{b10} I_\eta &\equiv &\int\; d{\bf v}_1 m_iV_{1,\gamma}V_{1,\beta}{\cal
K}_{ij,\gamma}[\partial_{V_\beta}f_j^{(0)}]\nonumber\\
&=&\chi_{ij}^{(0)}\sigma _{ij}^{d}m_i\int d{\bf v}_{1}\int d{\bf v}_{2}f_{i}( {\bf
V}_{1})\left(\partial_{V_{2,\beta}}
f_j^{(0)}({\bf V}_2)\right) \nonumber\\
& & \times\int d\widehat{\boldsymbol {\sigma }}\,\Theta (\widehat{\boldsymbol {\sigma}} \cdot {\bf
g})(\widehat{\boldsymbol {\sigma }}\cdot {\bf g})\widehat{\sigma}_\alpha \left(
V_{1,\gamma}'V_{1,\beta}'-V_{1,\gamma}V_{1,\beta}\right) \;,
\end{eqnarray}
where the identity (\ref{b1}) has been used. The scattering rule (\ref{a13}) gives
\begin{eqnarray}
\label{b11} V_{1,\gamma}'V_{1,\beta}'-V_{1,\gamma}V_{1,\beta}&=&-\mu_{ji}(1+ \alpha_{ij})(\widehat{\boldsymbol
{\sigma}}\cdot {\bf g})\left[ G_{ij,\gamma}\widehat{\sigma}_\beta+G_{ij,\beta}\widehat{\sigma}_\gamma+\mu_{ji}(
g_{\gamma}\widehat{\sigma}_\beta+g_{\beta}\widehat{\sigma}_\gamma\right.\nonumber\\
& & \left. -\mu_{ji}(1+\alpha_{ij})(\widehat{\boldsymbol{\sigma}}\cdot {\bf g})
\widehat{\sigma}_\gamma\widehat{\sigma}_\beta\right],
\end{eqnarray}
where ${\bf G}_{ij}=\mu_{ij}{\bf V}_1+\mu_{ji}{\bf V}_2$. Substitution of Eq.\ (\ref{b11}) into (\ref{b10})
allows the angular integral to be performed with the result
\begin{eqnarray}
\label{b12} \int d\widehat{\boldsymbol {\sigma }}\,\Theta (\widehat{\boldsymbol {\sigma}} \cdot {\bf
g})(\widehat{\boldsymbol {\sigma }}\cdot {\bf g})\widehat{\sigma}_\gamma \left(
V_{1,\gamma}'V_{1,\beta}'-V_{1,\gamma}V_{1,\beta}\right)&=&-\frac{B_2}{d+2}
\mu_{ji}(1+\alpha_{ij})\left[(d+3)g^2G_{ij,\beta}\right.\nonumber\\
&+& \left.\mu_{ji}(1+d-3\alpha_{ij})g^2g_\beta+2({\bf g}\cdot {\bf
G}_{ij})g_\beta \right].\nonumber\\
\end{eqnarray}
With this result the integral $I_\eta$ becomes
\begin{eqnarray}
\label{b14} I_\eta&=&-\frac{dB_2}{d+2}n_in_j\chi_{ij}^{(0)}\sigma_{ij}^{d}m_i\mu_{ji}(1+\alpha_{ij}) \int d{\bf
v}_{1}\int d{\bf v}_{2}f_{i}^{(0)}( {\bf V}_{1})f_{j}^{(0)}( {\bf
V}_{2})\nonumber\\
& &\times \left[(d+2)\mu_{ji}(3\alpha_{ij}-1)g^2-4(d+2)({\bf g}\cdot {\bf G}_{ij})
\right]\nonumber\\
&=&-dB_2n_in_j\chi_{ij}^{(0)}\sigma_{ij}^{d} m_i\mu_{ji}(1+\alpha_{ij})m_i\left[\mu_{ji}
\left(\frac{T_i}{m_i}+\frac{T_j}{m_j}\right)(3\alpha_{ij}-1)
-4\frac{T_i-T_j}{m_i+m_j}\right].\nonumber\\
\end{eqnarray}
From Eq.\ (\ref{b14}) it is easy to get the expression (\ref{3.7}).

To evaluate the collision integrals appearing in the determination of the heat flux one needs the partial
results
\begin{eqnarray}
\label{b15} {\bf S}_{i}({\bf V}_{1}^{^{\prime}})-{\bf S}_{i}({\bf V}_{1}) &=& \frac{m_{i}}{2}(1+\alpha
_{ij})\mu_{ji}(\widehat{\boldsymbol {\sigma}}\cdot {\bf g })\left\{ \left[
(1-\alpha_{ij}^{2})\mu_{ji}^{2}(\widehat{\boldsymbol {\sigma}}\cdot {\bf
g})^{2}-G_{ij}^{2}-\mu_{ji}^{2}g^{2}\right.
\right.   \nonumber \\
&&\left. -2\mu_{ji}({\bf g}\cdot {\bf G}_{ij})+2(1+\alpha_{ij})\mu _{ji}(\widehat{\boldsymbol {\sigma}}\cdot
{\bf g})(\widehat{\boldsymbol {\sigma }} \cdot {\bf G}_{ij})+(d+2)\frac{T_{i}}{m_{i}}\right]
\widehat{\boldsymbol {\sigma}}
\nonumber \\
&&\left.-\left[ (1-\alpha_{ij})\mu_{ji}(\widehat{\boldsymbol {\sigma }}\cdot {\bf g} )+2(\widehat{\boldsymbol
{\sigma}}\cdot {\bf G}_{ij})\right]\left( {\bf G}_{ij}+\mu_{ji}{\bf g}\right) \right\},
\end{eqnarray}
\begin{eqnarray}
\label{b16} \int d\widehat{\boldsymbol {\sigma}}\,\Theta (\widehat{\boldsymbol {\sigma}}\cdot {\bf g
})(\widehat{\boldsymbol {\sigma}}\cdot {\bf g})
 &&\widehat{\boldsymbol {\sigma}}\cdot\left[ {\bf S}_{i}( {\bf
V}_{1}^{^{\prime}})-{\bf S}_{i}({\bf V}_{1})\right] =-\frac{3}{2}\frac{B_2}{d+2}
(1+\alpha_{ij})\mu_{ji}\nonumber\\
& &\times \left\{ \frac{1}{3}\mu_{ji}^{2}\left[(d+2)-3\alpha_{ij}(1-\alpha_{ij})\right]
g^4+\frac{d+4}{3}g^2G_{ij}^2\right.\nonumber\\
& & \left.+\frac{1}{3}\mu_{ji}(7+2d-9\alpha_{ij})g^2 ({\bf g}\cdot {\bf G}_{ij})+\frac{4}{3}({\bf g}\cdot {\bf
G}_{ij})^2-\frac{(d+2)^2}{3}\frac{T_i}{m_i}g^2\right\}.\nonumber\\
\end{eqnarray}
The corresponding integrals associated with the heat flux can be explicitly evaluated by using Eqs.\ (\ref{b15})
and (\ref{b16}) and the same mathematical steps as before. After a lengthy algebra, one gets the expressions
(\ref{4.15a}), (\ref{4.17}), and (\ref{4.18}).

Let us consider now the integral appearing in the evaluation of the cooling rate. To do that, we compute first
the collision integral
\begin{equation}
\label{b17} I_\zeta=\int d{\bf v} V^4{\cal K}_{ij,\gamma} [
\partial_{V_{\gamma}}f_j^{(0)}].
\end{equation}
In order to evaluate it we use Eq.\ (\ref{b18}) and the angular integrations
\begin{eqnarray}
\label{b19}  \int d\widehat{\boldsymbol {\sigma }}\,\Theta (\widehat{\boldsymbol {\sigma}} \cdot {\bf
g})(\widehat{\boldsymbol {\sigma }}\cdot {\bf g})\widehat{\sigma}_\beta\,
\left(V_1'^4-V_1^4\right)&=&\frac{3B_2}{(d+2)(d+4)}
\mu_{ji}(1+\alpha_{ij})\nonumber\\
& & \times \left\{4\mu_{ji}(1+\alpha_{ij})\left[2({\bf V}_1\cdot {\bf g})^2g_\beta+2({\bf V}_1\cdot {\bf
g})g^2V_{1,\beta}\right.\right.\nonumber\\
& &\left. +\frac{1}{2}(d+6)g^2V_1^2g_\beta\right]
+5\mu_{ji}^3(1+\alpha_{ij})^3g^4g_\beta\nonumber\\
& &-\frac{4}{3}(d+4)V_1^2\left[2
({\bf V}_1\cdot {\bf g})g_{\beta}+g^2V_{1,\beta}\right]\nonumber\\
& & \left.-4\mu_{ji}^2(1+\alpha_{ij})^2g^2\left[4({\bf V}_1\cdot {\bf
g})g_{\beta}+g^2V_{1,\beta}\right]\right\}.
\end{eqnarray}
With these results, the integral $I_\zeta$ becomes
\begin{eqnarray}
\label{b20} I_{\zeta}&=&-3dB_2n_in_j\chi_{ij}^{(0)}\sigma_{ij}^{d}m_i\mu_{ji}(1+\alpha_{ij}) \int d{\bf
v}_{1}\int d{\bf v}_{2}f_{i}^{(0)}( {\bf V}_{1})f_{j}^{(0)}(
{\bf V}_{2})\nonumber\\
& \times& \left\{
 \mu_{ji}(1+\alpha_{ij})\left[2(d+4)g^2V_1^2+8({\bf V}_1\cdot{\bf
 g})^2
 +5\mu_{ji}^2(1+\alpha_{ij})^2g^4-16\mu_{ji}(1+\alpha_{ij})g^2
({\bf V}_1\cdot {\bf
g})\right]\right.\nonumber\\
& & \left.-\frac{8}{3}(d+2)V_1^2({\bf
V}_1\cdot {\bf g})\right\}\nonumber\\
&=&-3dB_2n_in_j\chi_{ij}^{(0)}\sigma_{ij}^d
\mu_{ji}(1+\alpha_{ij})\left\{\frac{T_i^2}{m_i^2}\left[\mu_{ji}(1+\alpha_{ij})\left(16+2d-16
\mu_{ji}(1+\alpha_{ij})\right.\right.\right.\nonumber\\
& &\left.\left. +5\mu_{ji}^2(1+\alpha_{ij})^2\right)-\frac{8}{3}(d+2)\right]+5\mu_{ji}^3
(1+\alpha_{ij})^3\frac{T_j^2}{m_j^2}
\nonumber\\
& & \left. +\frac{2}{d+2}\frac{T_iT_j}{m_im_j}
\mu_{ji}(1+\alpha_{ij})\left[d(d+4)+4-8(d+2)\mu_{ji}(1+\alpha_{ij})+5(d+2)
\mu_{ji}^2(1+\alpha_{ij})^2\right]\right\}.\nonumber\\
\end{eqnarray}
The final expression (\ref{1.16}) can be easily obtained from Eqs.\ (\ref{b14}) and (\ref{b20}).

Finally, note that the integrals $C_{ij}^T$ and $C_{ipj}^T$ defined by Eqs.\ (\ref{4.1e}) and (\ref{4.1f}),
respectively, can be easily computed by using the change of variables written below Eq.\ (\ref{a18}). After some
algebra and using the Maxwellian approach for the distributions $f_i^{(0)}$ one gets the results (\ref{4.1g})
and (\ref{4.1h}).

\section{Choice of $\chi_{ij}^{(0)}$ and $I_{i\ell j}$}
\label{appC}

This Appendix deals with the choice of the pair correlation function $\chi_{ij}^{(0)}$ and the functional
derivative $I_{i\ell j}$. A good approximation for $\chi_{ij}^{(0)}$ in two dimensions ($d=2$) is given by
\cite{SYM02}
\begin{equation}
\label{c1} \chi_{ij}^{(0)}=\frac{1}{1-\phi}+\frac{10-\phi}{16}\frac{\beta}{(1-\phi)^2}
\frac{\sigma_i\sigma_j}{\sigma_{ij}}-\frac{1}{16}\frac{\beta^2}{\phi(1-\phi)}
\left(\frac{\sigma_i\sigma_j}{\sigma_{ij}}\right)^2,
\end{equation}
where $\phi=\sum_{i=1}^s\phi_i$ is the total solid volume fraction, $\phi_i=n_i\pi \sigma_i^2/4$ is the species
volume fraction of component $i$ and $\beta=\pi(\sum_{i=1}^s n_i\sigma_i)/4$. In the case of hard-spheres
($d=3$) we take the following approximation for $\chi_{ij}^{(0)}$ \cite{Boublik70}
\begin{equation}
\label{c2} \chi_{ij}^{(0)}=\frac{1}{1-\phi}+\frac{3}{2}\frac{\beta}{(1-\phi)^2}
\frac{\sigma_i\sigma_j}{\sigma_{ij}}+\frac{1}{2}\frac{\beta^2}{(1-\phi)^3}
\left(\frac{\sigma_i\sigma_j}{\sigma_{ij}}\right)^2,
\end{equation}
where now $\phi_i=n_i\pi \sigma_i^3/6$ and $\beta=\pi(\sum_{i=1}^s n_i\sigma_i^2)/6$.

The parameter $I_{i\ell j}$ is chosen to recover the results derived by L\'opez de Haro {\em et al.}
\cite{deHaro83} for ordinary polydisperse mixtures in the context of the RET. To do that, for the sake of
simplicity, we assume that the temperature, the pressure and the flow velocity are homogeneous so that only the
spatial gradients associated with the partial densities will be considered. In this simple case, for elastic
collisions ($\alpha_{ij}=1$), the first-order distribution function is given by $f_i^{(1)}=\sum_{j=1}^s
\boldsymbol{\mathcal{B}}_{i}^j\cdot \nabla \ln n_j$ where $\boldsymbol{\mathcal{B}}_{i}^j({\bf V})$ verifies the
integral equation
\begin{equation}
\label{c3} \left(\boldsymbol{\mathcal{L}}{\cal B}^j\right)_{i,\gamma}=- V_\gamma
n_j\partial_{n_j}f_i^{(0)}-\sum_{\ell=1}^s\left( {\cal
K}_{i\ell,\gamma}[n_j\partial_{n_j}f_\ell^{(0)}]+\frac{1}{2}\left(n_\ell
\partial_{n_j}\ln \chi_{i\ell}^{(0)}+I_{i\ell j}\right){\cal
K}_{i\ell,\gamma}[f_\ell^{(0)}]\right).
\end{equation}
In the elastic case, $n_j\partial_{n_j}f_\ell^{(0)}=\delta_{j\ell}f_\ell^{(0)}$ and the linear operator
$\boldsymbol{\mathcal{K}}_{i\ell}[f_\ell^{(0)}]$ can be explicitly written as
\begin{equation}
\label{c4} {\cal K}_{i\ell,\gamma}[f_\ell^{(0)}]=2B_2n_\ell \chi_{i\ell}^{(0)}\sigma_{i\ell}^d V_\gamma
f_i^{(0)}({\bf V}).
\end{equation}
With this result, Eq.\ (\ref{c3}) becomes
\begin{equation}
\label{c5} -\left(\boldsymbol{\mathcal{L}}{\cal B}^j\right)_{i,\gamma}= V_\gamma
f_i^{(0)}\delta_{ij}+2B_2\sum_{\ell=1}^sn_\ell
\chi_{i\ell}^{(0)}\sigma_{i\ell}^d\left[\delta_{j\ell}+\frac{1}{2}\left(n_\ell
\partial_{n_j}\ln \chi_{i\ell}^{(0)}+I_{i\ell j}\right)\right]V_\gamma f_i^{(0)}.
\end{equation}
Comparison with the results derived by L\'opez de Haro {\em et al.} \cite{deHaro83} allow us to identify
$I_{ij\ell}$ to be defined through the relation
\begin{equation}
\label{c6} \sum_{\ell=1}^s n_\ell\chi_{i\ell}^{(0)}\sigma_{i\ell}^d \left(n_\ell
\partial_{n_j}\ln \chi_{i\ell}^{(0)}+I_{i\ell j}\right)=
\frac{n_j}{B_2}\left[\frac{1}{T}\left(\frac{\partial \mu_i}{\partial n_j}\right)_{T,n_{k\neq
j}}-\frac{1}{n_i}\delta_{ij}-2B_2\chi_{ij}^{(0)}\sigma_{ij}^d\right],
\end{equation}
where $\mu_i$ is the chemical potential of species $i$.

\bibliography{enskog}

\begin{thebibliography}{73}
\expandafter\ifx\csname natexlab\endcsname\relax\def\natexlab#1{#1}\fi
\expandafter\ifx\csname bibnamefont\endcsname\relax
  \def\bibnamefont#1{#1}\fi
\expandafter\ifx\csname bibfnamefont\endcsname\relax
  \def\bibfnamefont#1{#1}\fi
\expandafter\ifx\csname citenamefont\endcsname\relax
  \def\citenamefont#1{#1}\fi
\expandafter\ifx\csname url\endcsname\relax
  \def\url#1{\texttt{#1}}\fi
\expandafter\ifx\csname urlprefix\endcsname\relax\def\urlprefix{URL }\fi
\providecommand{\bibinfo}[2]{#2}
\providecommand{\eprint}[2][]{\url{#2}}

\bibitem[{\citenamefont{Garz\'o et~al.}(2007)\citenamefont{Garz\'o, Dufty, and
  Hrenya}}]{DGH06}
\bibinfo{author}{\bibfnamefont{V.}~\bibnamefont{Garz\'o}},
  \bibinfo{author}{\bibfnamefont{J.~W.} \bibnamefont{Dufty}}, \bibnamefont{and}
  \bibinfo{author}{\bibfnamefont{C.~M.} \bibnamefont{Hrenya}},
  \emph{\bibinfo{title}{Enskog theory for polydisperse granular mixtures {I}.
  {N}avier--{S}tokes order transport}}, \bibinfo{howpublished}{preprint
  cond-mat/0702109} (\bibinfo{year}{2007}).

\bibitem[{\citenamefont{Jenkins and Mancini}(1987)}]{Jenkins87}
\bibinfo{author}{\bibfnamefont{J.~T.} \bibnamefont{Jenkins}} \bibnamefont{and}
  \bibinfo{author}{\bibfnamefont{F.}~\bibnamefont{Mancini}},
  \bibinfo{journal}{J. Appl. Mech.-Trans. ASME} \textbf{\bibinfo{volume}{54}},
  \bibinfo{pages}{27} (\bibinfo{year}{1987}).

\bibitem[{\citenamefont{Jenkins and Mancini}(1989)}]{Jenkins89}
\bibinfo{author}{\bibfnamefont{J.~T.} \bibnamefont{Jenkins}} \bibnamefont{and}
  \bibinfo{author}{\bibfnamefont{F.}~\bibnamefont{Mancini}},
  \bibinfo{journal}{Phys. Fluids A} \textbf{\bibinfo{volume}{1}},
  \bibinfo{pages}{2050} (\bibinfo{year}{1989}).

\bibitem[{\citenamefont{Zamankhan}(1995)}]{Zamankhan95}
\bibinfo{author}{\bibfnamefont{P.}~\bibnamefont{Zamankhan}},
  \bibinfo{journal}{Phys. Rev. E} \textbf{\bibinfo{volume}{52}},
  \bibinfo{pages}{4877} (\bibinfo{year}{1995}).

\bibitem[{\citenamefont{Arnarson and Willits}(1998)}]{Arnarson98}
\bibinfo{author}{\bibfnamefont{B.~O.} \bibnamefont{Arnarson}} \bibnamefont{and}
  \bibinfo{author}{\bibfnamefont{J.~T.} \bibnamefont{Willits}},
  \bibinfo{journal}{Phys. Fluids} \textbf{\bibinfo{volume}{10}},
  \bibinfo{pages}{1324} (\bibinfo{year}{1998}).

\bibitem[{\citenamefont{Willits and Arnarson}(1999)}]{Willits99}
\bibinfo{author}{\bibfnamefont{J.~T.} \bibnamefont{Willits}} \bibnamefont{and}
  \bibinfo{author}{\bibfnamefont{B.~O.} \bibnamefont{Arnarson}},
  \bibinfo{journal}{Phys. Fluids} \textbf{\bibinfo{volume}{11}},
  \bibinfo{pages}{3116} (\bibinfo{year}{1999}).

\bibitem[{\citenamefont{Huilin et~al.}(2001)\citenamefont{Huilin, Gidaspow, and
  Manger}}]{Huilin01}
\bibinfo{author}{\bibfnamefont{L.}~\bibnamefont{Huilin}},
  \bibinfo{author}{\bibfnamefont{D.}~\bibnamefont{Gidaspow}}, \bibnamefont{and}
  \bibinfo{author}{\bibfnamefont{E.}~\bibnamefont{Manger}},
  \bibinfo{journal}{Phys. Rev. E} \textbf{\bibinfo{volume}{64}},
  \bibinfo{pages}{061301} (\bibinfo{year}{2001}).

\bibitem[{\citenamefont{Rahaman et~al.}(2003)\citenamefont{Rahaman, Naser, and
  Witt}}]{Rahaman03}
\bibinfo{author}{\bibfnamefont{M.~F.} \bibnamefont{Rahaman}},
  \bibinfo{author}{\bibfnamefont{J.}~\bibnamefont{Naser}}, \bibnamefont{and}
  \bibinfo{author}{\bibfnamefont{P.~J.} \bibnamefont{Witt}},
  \bibinfo{journal}{Powder Technol.} \textbf{\bibinfo{volume}{138}},
  \bibinfo{pages}{82} (\bibinfo{year}{2003}).

\bibitem[{\citenamefont{Iddir and Arastoopour}(2005)}]{Iddir05}
\bibinfo{author}{\bibfnamefont{H.}~\bibnamefont{Iddir}} \bibnamefont{and}
  \bibinfo{author}{\bibfnamefont{H.}~\bibnamefont{Arastoopour}},
  \bibinfo{journal}{AIChE J.} \textbf{\bibinfo{volume}{51}},
  \bibinfo{pages}{1620} (\bibinfo{year}{2005}).

\bibitem[{\citenamefont{Serero et~al.}(2006)\citenamefont{Serero, Goldhirsch,
  Noskowicz, and Tan}}]{Serero06}
\bibinfo{author}{\bibfnamefont{D.}~\bibnamefont{Serero}},
  \bibinfo{author}{\bibfnamefont{I.}~\bibnamefont{Goldhirsch}},
  \bibinfo{author}{\bibfnamefont{S.~H.} \bibnamefont{Noskowicz}},
  \bibnamefont{and} \bibinfo{author}{\bibfnamefont{M.~L.} \bibnamefont{Tan}},
  \bibinfo{journal}{J. Fluid Mech.} \textbf{\bibinfo{volume}{554}},
  \bibinfo{pages}{237} (\bibinfo{year}{2006}).

\bibitem[{\citenamefont{Garz\'o and Dufty}(2002)}]{Garzo02}
\bibinfo{author}{\bibfnamefont{V.}~\bibnamefont{Garz\'o}} \bibnamefont{and}
  \bibinfo{author}{\bibfnamefont{J.~W.} \bibnamefont{Dufty}},
  \bibinfo{journal}{Phys. Fluids} \textbf{\bibinfo{volume}{14}},
  \bibinfo{pages}{1476} (\bibinfo{year}{2002}).

\bibitem[{\citenamefont{Garz\'o and Dufty}(1999{\natexlab{a}})}]{Garzo99}
\bibinfo{author}{\bibfnamefont{V.}~\bibnamefont{Garz\'o}} \bibnamefont{and}
  \bibinfo{author}{\bibfnamefont{J.~W.} \bibnamefont{Dufty}},
  \bibinfo{journal}{Phys. Rev. E} \textbf{\bibinfo{volume}{60}},
  \bibinfo{pages}{5706} (\bibinfo{year}{1999}{\natexlab{a}}).

\bibitem[{\citenamefont{Garz\'o et~al.}(2006)\citenamefont{Garz\'o, Montanero,
  and Dufty}}]{GMD06}
\bibinfo{author}{\bibfnamefont{V.}~\bibnamefont{Garz\'o}},
  \bibinfo{author}{\bibfnamefont{J.~M.} \bibnamefont{Montanero}},
  \bibnamefont{and} \bibinfo{author}{\bibfnamefont{J.~W.} \bibnamefont{Dufty}},
  \bibinfo{journal}{Phys. Fluids} \textbf{\bibinfo{volume}{18}},
  \bibinfo{pages}{083305} (\bibinfo{year}{2006}).

\bibitem[{\citenamefont{Garz\'o and Dufty}(1999{\natexlab{b}})}]{Garzo99a}
\bibinfo{author}{\bibfnamefont{V.}~\bibnamefont{Garz\'o}} \bibnamefont{and}
  \bibinfo{author}{\bibfnamefont{J.~W.} \bibnamefont{Dufty}},
  \bibinfo{journal}{Phys. Rev. E} \textbf{\bibinfo{volume}{59}},
  \bibinfo{pages}{5895} (\bibinfo{year}{1999}{\natexlab{b}}).

\bibitem[{\citenamefont{Lutsko}(2005)}]{Lutsko05}
\bibinfo{author}{\bibfnamefont{J.~F.} \bibnamefont{Lutsko}},
  \bibinfo{journal}{Phys. Rev. E} \textbf{\bibinfo{volume}{72}},
  \bibinfo{pages}{021306} (\bibinfo{year}{2005}).

\bibitem[{\citenamefont{Garz\'o and Montanero}()}]{GM06}
\bibinfo{author}{\bibfnamefont{V.}~\bibnamefont{Garz\'o}} \bibnamefont{and}
  \bibinfo{author}{\bibfnamefont{J.~M.} \bibnamefont{Montanero}},
  \bibinfo{howpublished}{J. Stat. Phys. (in press) and preprint
  cond-mat/0604552}.

\bibitem[{\citenamefont{Garz\'o and Montanero}(2003)}]{GM03}
\bibinfo{author}{\bibfnamefont{V.}~\bibnamefont{Garz\'o}} \bibnamefont{and}
  \bibinfo{author}{\bibfnamefont{J.~M.} \bibnamefont{Montanero}},
  \bibinfo{journal}{Phys. Rev. E} \textbf{\bibinfo{volume}{68}},
  \bibinfo{pages}{041302} (\bibinfo{year}{2003}).

\bibitem[{\citenamefont{Barajas et~al.}(1973)\citenamefont{Barajas,
  Garcia-Col\'{\i}n, and Pi\~{n}a}}]{Barajas73}
\bibinfo{author}{\bibfnamefont{L.}~\bibnamefont{Barajas}},
  \bibinfo{author}{\bibfnamefont{L.~S.} \bibnamefont{Garcia-Col\'{\i}n}},
  \bibnamefont{and} \bibinfo{author}{\bibfnamefont{E.}~\bibnamefont{Pi\~{n}a}},
  \bibinfo{journal}{J. Stat. Phys.} \textbf{\bibinfo{volume}{7}},
  \bibinfo{pages}{161} (\bibinfo{year}{1973}).

\bibitem[{\citenamefont{van Beijeren and
  Ernst}(1973{\natexlab{a}})}]{vanBeijeren73}
\bibinfo{author}{\bibfnamefont{H.}~\bibnamefont{van Beijeren}}
  \bibnamefont{and} \bibinfo{author}{\bibfnamefont{M.~H.} \bibnamefont{Ernst}},
  \bibinfo{journal}{Physica} \textbf{\bibinfo{volume}{68}},
  \bibinfo{pages}{437} (\bibinfo{year}{1973}{\natexlab{a}}).

\bibitem[{\citenamefont{van Beijeren and
  Ernst}(1973{\natexlab{b}})}]{vanBeijeren73a}
\bibinfo{author}{\bibfnamefont{H.}~\bibnamefont{van Beijeren}}
  \bibnamefont{and} \bibinfo{author}{\bibfnamefont{M.~H.} \bibnamefont{Ernst}},
  \bibinfo{journal}{Physica} \textbf{\bibinfo{volume}{70}},
  \bibinfo{pages}{225} (\bibinfo{year}{1973}{\natexlab{b}}).

\bibitem[{\citenamefont{van Beijeren and Ernst}(1979)}]{vanBeijeren79}
\bibinfo{author}{\bibfnamefont{H.}~\bibnamefont{van Beijeren}}
  \bibnamefont{and} \bibinfo{author}{\bibfnamefont{M.~H.} \bibnamefont{Ernst}},
  \bibinfo{journal}{J. Stat. Phys.} \textbf{\bibinfo{volume}{21}},
  \bibinfo{pages}{125} (\bibinfo{year}{1979}).

\bibitem[{\citenamefont{L\'opez~de Haro and Garz\'o}(1993)}]{LG93}
\bibinfo{author}{\bibfnamefont{M.}~\bibnamefont{L\'opez~de Haro}}
  \bibnamefont{and} \bibinfo{author}{\bibfnamefont{V.}~\bibnamefont{Garz\'o}},
  \bibinfo{journal}{Physica A} \textbf{\bibinfo{volume}{197}},
  \bibinfo{pages}{98} (\bibinfo{year}{1993}).

\bibitem[{\citenamefont{Arnarson and Jenkins}(2000)}]{Arnarson00}
\bibinfo{author}{\bibfnamefont{B.~O.} \bibnamefont{Arnarson}} \bibnamefont{and}
  \bibinfo{author}{\bibfnamefont{J.~T.} \bibnamefont{Jenkins}}, in
  \emph{\bibinfo{booktitle}{Traffic and granular flow '99: Social, traffic, and
  granular dynamics}}, edited by
  \bibinfo{editor}{\bibfnamefont{D.}~\bibnamefont{Helging}},
  \bibinfo{editor}{\bibfnamefont{H.~J.} \bibnamefont{Herrmann}},
  \bibinfo{editor}{\bibfnamefont{M.}~\bibnamefont{Schreckenberg}},
  \bibnamefont{and} \bibinfo{editor}{\bibfnamefont{D.~E.} \bibnamefont{Wolf}}
  (\bibinfo{publisher}{Springer}, \bibinfo{year}{2000}).

\bibitem[{\citenamefont{Sela and Goldhirsch}(1998)}]{Sela98}
\bibinfo{author}{\bibfnamefont{N.}~\bibnamefont{Sela}} \bibnamefont{and}
  \bibinfo{author}{\bibfnamefont{I.}~\bibnamefont{Goldhirsch}},
  \bibinfo{journal}{J. Fluid Mech.} \textbf{\bibinfo{volume}{361}},
  \bibinfo{pages}{41} (\bibinfo{year}{1998}).

\bibitem[{\citenamefont{Brey et~al.}(2001)\citenamefont{Brey, Ruiz-Montero, and
  Moreno}}]{Brey01}
\bibinfo{author}{\bibfnamefont{J.~J.} \bibnamefont{Brey}},
  \bibinfo{author}{\bibfnamefont{M.~J.} \bibnamefont{Ruiz-Montero}},
  \bibnamefont{and} \bibinfo{author}{\bibfnamefont{F.}~\bibnamefont{Moreno}},
  \bibinfo{journal}{Phys. Rev. E} \textbf{\bibinfo{volume}{63}},
  \bibinfo{pages}{061305} (\bibinfo{year}{2001}).

\bibitem[{\citenamefont{Martin et~al.}(2005)\citenamefont{Martin, Huntley, and
  Wildman}}]{Martin05}
\bibinfo{author}{\bibfnamefont{T.~W.} \bibnamefont{Martin}},
  \bibinfo{author}{\bibfnamefont{J.~M.} \bibnamefont{Huntley}},
  \bibnamefont{and} \bibinfo{author}{\bibfnamefont{R.~D.}
  \bibnamefont{Wildman}}, \bibinfo{journal}{J. Fluid Mech.}
  \textbf{\bibinfo{volume}{535}}, \bibinfo{pages}{325} (\bibinfo{year}{2005}).

\bibitem[{\citenamefont{Behringer et~al.}(2002)\citenamefont{Behringer, van
  Doorn, Hartley, and Pak}}]{Behringer02}
\bibinfo{author}{\bibfnamefont{R.~P.} \bibnamefont{Behringer}},
  \bibinfo{author}{\bibfnamefont{E.}~\bibnamefont{van Doorn}},
  \bibinfo{author}{\bibfnamefont{R.~R.} \bibnamefont{Hartley}},
  \bibnamefont{and} \bibinfo{author}{\bibfnamefont{H.~K.} \bibnamefont{Pak}},
  \bibinfo{journal}{Gran. Matt.} \textbf{\bibinfo{volume}{4}},
  \bibinfo{pages}{9} (\bibinfo{year}{2002}).

\bibitem[{\citenamefont{Wassgren et~al.}(2003)\citenamefont{Wassgren, Cordova,
  Zenit, and Karion}}]{Wassgren03}
\bibinfo{author}{\bibfnamefont{C.~R.} \bibnamefont{Wassgren}},
  \bibinfo{author}{\bibfnamefont{J.~A.} \bibnamefont{Cordova}},
  \bibinfo{author}{\bibfnamefont{R.}~\bibnamefont{Zenit}}, \bibnamefont{and}
  \bibinfo{author}{\bibfnamefont{A.}~\bibnamefont{Karion}},
  \bibinfo{journal}{Phys. Fluids} \textbf{\bibinfo{volume}{15}},
  \bibinfo{pages}{3318} (\bibinfo{year}{2003}).

\bibitem[{\citenamefont{Santos et~al.}(2004)\citenamefont{Santos, Garz\'o, and
  Dufty}}]{Santos04}
\bibinfo{author}{\bibfnamefont{A.}~\bibnamefont{Santos}},
  \bibinfo{author}{\bibfnamefont{V.}~\bibnamefont{Garz\'o}}, \bibnamefont{and}
  \bibinfo{author}{\bibfnamefont{J.~W.} \bibnamefont{Dufty}},
  \bibinfo{journal}{Phys. Rev. E} \textbf{\bibinfo{volume}{69}},
  \bibinfo{pages}{061303} (\bibinfo{year}{2004}).

\bibitem[{\citenamefont{Galvin et~al.}(in press)\citenamefont{Galvin, Hrenya,
  and Wildman}}]{Galvin07}
\bibinfo{author}{\bibfnamefont{J.~E.} \bibnamefont{Galvin}},
  \bibinfo{author}{\bibfnamefont{C.~M.} \bibnamefont{Hrenya}},
  \bibnamefont{and} \bibinfo{author}{\bibfnamefont{R.~D.}
  \bibnamefont{Wildman}}, \bibinfo{journal}{J. Fluid Mech.}  (\bibinfo{year}{in
  press}).

\bibitem[{\citenamefont{Hrenya et~al.}(in preparation)\citenamefont{Hrenya,
  Galvin, and Wildman}}]{Hrenya06}
\bibinfo{author}{\bibfnamefont{C.~M.} \bibnamefont{Hrenya}},
  \bibinfo{author}{\bibfnamefont{J.~E.} \bibnamefont{Galvin}},
  \bibnamefont{and} \bibinfo{author}{\bibfnamefont{R.~D.}
  \bibnamefont{Wildman}} (\bibinfo{year}{in preparation}).

\bibitem[{\citenamefont{Kumaran}(1997)}]{Kumaran97}
\bibinfo{author}{\bibfnamefont{V.}~\bibnamefont{Kumaran}}, \bibinfo{journal}{J.
  Fluid Mech.} \textbf{\bibinfo{volume}{340}}, \bibinfo{pages}{319}
  (\bibinfo{year}{1997}).

\bibitem[{\citenamefont{Risso and Cordero}(2002)}]{Risso02}
\bibinfo{author}{\bibfnamefont{D.}~\bibnamefont{Risso}} \bibnamefont{and}
  \bibinfo{author}{\bibfnamefont{P.}~\bibnamefont{Cordero}},
  \bibinfo{journal}{Phys. Rev. E} \textbf{\bibinfo{volume}{65}},
  \bibinfo{pages}{021304} (\bibinfo{year}{2002}).

\bibitem[{\citenamefont{Kumaran}(2005)}]{Kumaran05}
\bibinfo{author}{\bibfnamefont{V.}~\bibnamefont{Kumaran}},
  \bibinfo{journal}{Phys. Rev. Lett.} \textbf{\bibinfo{volume}{95}},
  \bibinfo{pages}{108001} (\bibinfo{year}{2005}).

\bibitem[{\citenamefont{Goldhirsch}(1999)}]{Goldhirsch99}
\bibinfo{author}{\bibfnamefont{I.}~\bibnamefont{Goldhirsch}},
  \bibinfo{journal}{Chaos} \textbf{\bibinfo{volume}{9}}, \bibinfo{pages}{659}
  (\bibinfo{year}{1999}).

\bibitem[{\citenamefont{Montanero and Garz\'o}(2002{\natexlab{a}})}]{MG02}
\bibinfo{author}{\bibfnamefont{J.~M.} \bibnamefont{Montanero}}
  \bibnamefont{and} \bibinfo{author}{\bibfnamefont{V.}~\bibnamefont{Garz\'o}},
  \bibinfo{journal}{Physica A} \textbf{\bibinfo{volume}{310}},
  \bibinfo{pages}{17} (\bibinfo{year}{2002}{\natexlab{a}}).

\bibitem[{\citenamefont{Lutsko}(2004)}]{L04}
\bibinfo{author}{\bibfnamefont{J.~F.} \bibnamefont{Lutsko}},
  \bibinfo{journal}{Phys. Rev. E} \textbf{\bibinfo{volume}{70}},
  \bibinfo{pages}{061101} (\bibinfo{year}{2004}).

\bibitem[{\citenamefont{Jenkins}(1998)}]{Jenkins98}
\bibinfo{author}{\bibfnamefont{J.~T.} \bibnamefont{Jenkins}}, in
  \emph{\bibinfo{booktitle}{Physics of dry granular media}}, edited by
  \bibinfo{editor}{\bibfnamefont{H.~J.} \bibnamefont{Hermann}},
  \bibinfo{editor}{\bibfnamefont{J.~P.} \bibnamefont{Hovi}}, \bibnamefont{and}
  \bibinfo{editor}{\bibfnamefont{S.}~\bibnamefont{Luding}}
  (\bibinfo{publisher}{Kluwer}, \bibinfo{year}{1998}).

\bibitem[{\citenamefont{Montanero and Garz\'o}(2003)}]{Montanero03}
\bibinfo{author}{\bibfnamefont{J.~M.} \bibnamefont{Montanero}}
  \bibnamefont{and} \bibinfo{author}{\bibfnamefont{V.}~\bibnamefont{Garz\'o}},
  \bibinfo{journal}{Phys. Rev. E} \textbf{\bibinfo{volume}{67}},
  \bibinfo{pages}{021308} (\bibinfo{year}{2003}).

\bibitem[{\citenamefont{Garz\'o and Montanero}(2004)}]{Garzo04}
\bibinfo{author}{\bibfnamefont{V.}~\bibnamefont{Garz\'o}} \bibnamefont{and}
  \bibinfo{author}{\bibfnamefont{J.~M.} \bibnamefont{Montanero}},
  \bibinfo{journal}{Phys. Rev. E} \textbf{\bibinfo{volume}{69}},
  \bibinfo{pages}{021301} (\bibinfo{year}{2004}).

\bibitem[{\citenamefont{Chapman and Cowling}(1970)}]{Chapman70}
\bibinfo{author}{\bibfnamefont{S.}~\bibnamefont{Chapman}} \bibnamefont{and}
  \bibinfo{author}{\bibfnamefont{T.~G.} \bibnamefont{Cowling}},
  \emph{\bibinfo{title}{The Mathematical Theory of Non-Uniform Gases}}
  (\bibinfo{publisher}{Cambridge University Press},
  \bibinfo{address}{Cambridge}, \bibinfo{year}{1970}).

\bibitem[{\citenamefont{Campbell}(1990)}]{Campbell90}
\bibinfo{author}{\bibfnamefont{C.~S.} \bibnamefont{Campbell}},
  \bibinfo{journal}{Annu. Rev. Fluid Mech.} \textbf{\bibinfo{volume}{22}},
  \bibinfo{pages}{57} (\bibinfo{year}{1990}).

\bibitem[{\citenamefont{Goldshtein and Shapiro}(1995)}]{Goldshtein95}
\bibinfo{author}{\bibfnamefont{A.}~\bibnamefont{Goldshtein}} \bibnamefont{and}
  \bibinfo{author}{\bibfnamefont{M.}~\bibnamefont{Shapiro}},
  \bibinfo{journal}{J. Fluid Mech.} \textbf{\bibinfo{volume}{282}},
  \bibinfo{pages}{75} (\bibinfo{year}{1995}).

\bibitem[{\citenamefont{Goldhirsch and Tan}(1996)}]{Goldhirsch96}
\bibinfo{author}{\bibfnamefont{I.}~\bibnamefont{Goldhirsch}} \bibnamefont{and}
  \bibinfo{author}{\bibfnamefont{M.~L.} \bibnamefont{Tan}},
  \bibinfo{journal}{Phys. Fluids} \textbf{\bibinfo{volume}{8}},
  \bibinfo{pages}{1752} (\bibinfo{year}{1996}).

\bibitem[{\citenamefont{Esipov and P\"oschel}(1997)}]{Esipov97}
\bibinfo{author}{\bibfnamefont{S.~E.} \bibnamefont{Esipov}} \bibnamefont{and}
  \bibinfo{author}{\bibfnamefont{T.}~\bibnamefont{P\"oschel}},
  \bibinfo{journal}{J. Stat. Phys.} \textbf{\bibinfo{volume}{86}},
  \bibinfo{pages}{1385} (\bibinfo{year}{1997}).

\bibitem[{\citenamefont{van Noije and Ernst}(1998)}]{vanNoije98}
\bibinfo{author}{\bibfnamefont{T.~P.~C.} \bibnamefont{van Noije}}
  \bibnamefont{and} \bibinfo{author}{\bibfnamefont{M.~H.} \bibnamefont{Ernst}},
  \bibinfo{journal}{Gran. Matt.} \textbf{\bibinfo{volume}{1}},
  \bibinfo{pages}{57} (\bibinfo{year}{1998}).

\bibitem[{\citenamefont{Brey et~al.}(1999)\citenamefont{Brey, Cubero, and
  Ruiz-Montero}}]{Brey99}
\bibinfo{author}{\bibfnamefont{J.~J.} \bibnamefont{Brey}},
  \bibinfo{author}{\bibfnamefont{D.}~\bibnamefont{Cubero}}, \bibnamefont{and}
  \bibinfo{author}{\bibfnamefont{M.~J.} \bibnamefont{Ruiz-Montero}},
  \bibinfo{journal}{Phys. Rev. E} \textbf{\bibinfo{volume}{59}},
  \bibinfo{pages}{1256} (\bibinfo{year}{1999}).

\bibitem[{\citenamefont{Losert et~al.}(1999)\citenamefont{Losert, Cooper,
  Delour, Kudrolli, and Gollub}}]{Losert99}
\bibinfo{author}{\bibfnamefont{W.}~\bibnamefont{Losert}},
  \bibinfo{author}{\bibfnamefont{D.~G.~W.} \bibnamefont{Cooper}},
  \bibinfo{author}{\bibfnamefont{J.}~\bibnamefont{Delour}},
  \bibinfo{author}{\bibfnamefont{A.}~\bibnamefont{Kudrolli}}, \bibnamefont{and}
  \bibinfo{author}{\bibfnamefont{J.~P.} \bibnamefont{Gollub}},
  \bibinfo{journal}{Chaos} \textbf{\bibinfo{volume}{9}}, \bibinfo{pages}{682}
  (\bibinfo{year}{1999}).

\bibitem[{\citenamefont{Kudrolli and Henry}(2000)}]{Kudrolli00}
\bibinfo{author}{\bibfnamefont{A.}~\bibnamefont{Kudrolli}} \bibnamefont{and}
  \bibinfo{author}{\bibfnamefont{J.}~\bibnamefont{Henry}},
  \bibinfo{journal}{Phys. Rev. E} \textbf{\bibinfo{volume}{62}},
  \bibinfo{pages}{R1489} (\bibinfo{year}{2000}).

\bibitem[{\citenamefont{Barrat and Trizac}(2002)}]{Barrat02}
\bibinfo{author}{\bibfnamefont{A.}~\bibnamefont{Barrat}} \bibnamefont{and}
  \bibinfo{author}{\bibfnamefont{E.}~\bibnamefont{Trizac}},
  \bibinfo{journal}{Gran. Matt.} \textbf{\bibinfo{volume}{4}},
  \bibinfo{pages}{57} (\bibinfo{year}{2002}).

\bibitem[{\citenamefont{Clelland and Hrenya}(2002)}]{Clelland02}
\bibinfo{author}{\bibfnamefont{R.}~\bibnamefont{Clelland}} \bibnamefont{and}
  \bibinfo{author}{\bibfnamefont{C.~M.} \bibnamefont{Hrenya}},
  \bibinfo{journal}{Phys. Rev. E} \textbf{\bibinfo{volume}{65}},
  \bibinfo{pages}{031301} (\bibinfo{year}{2002}).

\bibitem[{\citenamefont{Dahl et~al.}(2002{\natexlab{a}})\citenamefont{Dahl,
  Clelland, and Hrenya}}]{Dahl02}
\bibinfo{author}{\bibfnamefont{S.~R.} \bibnamefont{Dahl}},
  \bibinfo{author}{\bibfnamefont{R.}~\bibnamefont{Clelland}}, \bibnamefont{and}
  \bibinfo{author}{\bibfnamefont{C.~M.} \bibnamefont{Hrenya}},
  \bibinfo{journal}{Phys. Fluids} \textbf{\bibinfo{volume}{14}},
  \bibinfo{pages}{1972} (\bibinfo{year}{2002}{\natexlab{a}}).

\bibitem[{\citenamefont{Montanero and Garz\'o}(2002{\natexlab{b}})}]{MG02bis}
\bibinfo{author}{\bibfnamefont{J.~M.} \bibnamefont{Montanero}}
  \bibnamefont{and} \bibinfo{author}{\bibfnamefont{V.}~\bibnamefont{Garz\'o}},
  \bibinfo{journal}{Gran. Matt.} \textbf{\bibinfo{volume}{4}},
  \bibinfo{pages}{17} (\bibinfo{year}{2002}{\natexlab{b}}).

\bibitem[{\citenamefont{Dahl et~al.}(2002{\natexlab{b}})\citenamefont{Dahl,
  Hrenya, Garz\'o, and Dufty}}]{Dahl2-02}
\bibinfo{author}{\bibfnamefont{S.~R.} \bibnamefont{Dahl}},
  \bibinfo{author}{\bibfnamefont{C.~M.} \bibnamefont{Hrenya}},
  \bibinfo{author}{\bibfnamefont{V.}~\bibnamefont{Garz\'o}}, \bibnamefont{and}
  \bibinfo{author}{\bibfnamefont{J.~W.} \bibnamefont{Dufty}},
  \bibinfo{journal}{Phys. Rev. E} \textbf{\bibinfo{volume}{66}},
  \bibinfo{pages}{041301} (\bibinfo{year}{2002}{\natexlab{b}}).

\bibitem[{\citenamefont{Alam and Luding}(2003)}]{Alam03}
\bibinfo{author}{\bibfnamefont{M.}~\bibnamefont{Alam}} \bibnamefont{and}
  \bibinfo{author}{\bibfnamefont{S.}~\bibnamefont{Luding}},
  \bibinfo{journal}{J. Fluid Mech.} \textbf{\bibinfo{volume}{476}},
  \bibinfo{pages}{69} (\bibinfo{year}{2003}).

\bibitem[{\citenamefont{Paolotti et~al.}(2003)\citenamefont{Paolotti, Cattuto,
  Marconi, and Puglisi}}]{Paolotti03}
\bibinfo{author}{\bibfnamefont{D.}~\bibnamefont{Paolotti}},
  \bibinfo{author}{\bibfnamefont{C.}~\bibnamefont{Cattuto}},
  \bibinfo{author}{\bibfnamefont{U.~M.~B.} \bibnamefont{Marconi}},
  \bibnamefont{and} \bibinfo{author}{\bibfnamefont{A.}~\bibnamefont{Puglisi}},
  \bibinfo{journal}{Gran. Matt.} \textbf{\bibinfo{volume}{5}},
  \bibinfo{pages}{75} (\bibinfo{year}{2003}).

\bibitem[{\citenamefont{Feitosa and Menon}(2002)}]{Feitosa02}
\bibinfo{author}{\bibfnamefont{K.}~\bibnamefont{Feitosa}} \bibnamefont{and}
  \bibinfo{author}{\bibfnamefont{N.}~\bibnamefont{Menon}},
  \bibinfo{journal}{Phys. Rev. Lett.} \textbf{\bibinfo{volume}{88}},
  \bibinfo{pages}{198301} (\bibinfo{year}{2002}).

\bibitem[{\citenamefont{Wildman and Parker}(2002)}]{Wildman02}
\bibinfo{author}{\bibfnamefont{R.~D.} \bibnamefont{Wildman}} \bibnamefont{and}
  \bibinfo{author}{\bibfnamefont{D.~J.} \bibnamefont{Parker}},
  \bibinfo{journal}{Phys. Rev. Lett.} \textbf{\bibinfo{volume}{88}},
  \bibinfo{pages}{064301} (\bibinfo{year}{2002}).

\bibitem[{\citenamefont{Schr\"oter et~al.}(2006)\citenamefont{Schr\"oter,
  Ulrich, Kreft, Swift, and Swinney}}]{SUKSS06}
\bibinfo{author}{\bibfnamefont{M.}~\bibnamefont{Schr\"oter}},
  \bibinfo{author}{\bibfnamefont{S.}~\bibnamefont{Ulrich}},
  \bibinfo{author}{\bibfnamefont{J.}~\bibnamefont{Kreft}},
  \bibinfo{author}{\bibfnamefont{J.~B.} \bibnamefont{Swift}}, \bibnamefont{and}
  \bibinfo{author}{\bibfnamefont{H.~L.} \bibnamefont{Swinney}},
  \bibinfo{journal}{Phys. Rev. E} \textbf{\bibinfo{volume}{74}},
  \bibinfo{pages}{011307} (\bibinfo{year}{2006}).

\bibitem[{\citenamefont{Galvin et~al.}(2005)\citenamefont{Galvin, Dahl, and
  Hrenya}}]{Galvin05}
\bibinfo{author}{\bibfnamefont{J.~E.} \bibnamefont{Galvin}},
  \bibinfo{author}{\bibfnamefont{S.~R.} \bibnamefont{Dahl}}, \bibnamefont{and}
  \bibinfo{author}{\bibfnamefont{C.~M.} \bibnamefont{Hrenya}},
  \bibinfo{journal}{J. Fluid Mech.} \textbf{\bibinfo{volume}{528}},
  \bibinfo{pages}{207} (\bibinfo{year}{2005}).

\bibitem[{\citenamefont{Brey et~al.}(2005)\citenamefont{Brey, Ruiz-Montero, and
  Moreno}}]{Brey05}
\bibinfo{author}{\bibfnamefont{J.~J.} \bibnamefont{Brey}},
  \bibinfo{author}{\bibfnamefont{M.~J.} \bibnamefont{Ruiz-Montero}},
  \bibnamefont{and} \bibinfo{author}{\bibfnamefont{F.}~\bibnamefont{Moreno}},
  \bibinfo{journal}{Phys. Rev. Lett.} \textbf{\bibinfo{volume}{95}},
  \bibinfo{pages}{0978001} (\bibinfo{year}{2005}).

\bibitem[{\citenamefont{Garz\'o}(2006)}]{G06}
\bibinfo{author}{\bibfnamefont{V.}~\bibnamefont{Garz\'o}},
  \bibinfo{journal}{Europhys. Lett.} \textbf{\bibinfo{volume}{75}},
  \bibinfo{pages}{521} (\bibinfo{year}{2006}).

\bibitem[{\citenamefont{Yoon and Jenkins}(2006)}]{Yoon06}
\bibinfo{author}{\bibfnamefont{D.~K.} \bibnamefont{Yoon}} \bibnamefont{and}
  \bibinfo{author}{\bibfnamefont{J.~T.} \bibnamefont{Jenkins}},
  \bibinfo{journal}{Phys. Fluids} \textbf{\bibinfo{volume}{18}},
  \bibinfo{pages}{073303} (\bibinfo{year}{2006}).

\bibitem[{\citenamefont{Ferziger and Kaper}(1972)}]{Ferziger72}
\bibinfo{author}{\bibfnamefont{J.}~\bibnamefont{Ferziger}} \bibnamefont{and}
  \bibinfo{author}{\bibfnamefont{H.}~\bibnamefont{Kaper}},
  \emph{\bibinfo{title}{Mathematical Theory of Transport Processes in Gases}}
  (\bibinfo{publisher}{North Holland}, \bibinfo{address}{Amsterdam},
  \bibinfo{year}{1972}).

\bibitem[{\citenamefont{Mansoori et~al.}(1971)\citenamefont{Mansoori, Carnahan,
  Starling, and Leland}}]{Mansoori71}
\bibinfo{author}{\bibfnamefont{G.~A.} \bibnamefont{Mansoori}},
  \bibinfo{author}{\bibfnamefont{N.~F.} \bibnamefont{Carnahan}},
  \bibinfo{author}{\bibfnamefont{K.~E.} \bibnamefont{Starling}},
  \bibnamefont{and} \bibinfo{author}{\bibfnamefont{T.~W.}
  \bibnamefont{Leland}}, \bibinfo{journal}{J. Chem. Phys.}
  \textbf{\bibinfo{volume}{54}}, \bibinfo{pages}{1523} (\bibinfo{year}{1971}).

\bibitem[{\citenamefont{Carnahan and Starling}(1969)}]{Carnahan69}
\bibinfo{author}{\bibfnamefont{N.~F.} \bibnamefont{Carnahan}} \bibnamefont{and}
  \bibinfo{author}{\bibfnamefont{K.~E.} \bibnamefont{Starling}},
  \bibinfo{journal}{J. Chem. Phys.} \textbf{\bibinfo{volume}{51}},
  \bibinfo{pages}{635} (\bibinfo{year}{1969}).

\bibitem[{\citenamefont{Alam et~al.}(2002)\citenamefont{Alam, Willits,
  Arnarson, and Luding}}]{Alam02}
\bibinfo{author}{\bibfnamefont{M.}~\bibnamefont{Alam}},
  \bibinfo{author}{\bibfnamefont{J.}~\bibnamefont{Willits}},
  \bibinfo{author}{\bibfnamefont{B.}~\bibnamefont{Arnarson}}, \bibnamefont{and}
  \bibinfo{author}{\bibfnamefont{S.}~\bibnamefont{Luding}},
  \bibinfo{journal}{Phys. Fluids} \textbf{\bibinfo{volume}{11}},
  \bibinfo{pages}{4085} (\bibinfo{year}{2002}).

\bibitem[{\citenamefont{Mathiesen et~al.}(2000)\citenamefont{Mathiesen,
  Solberg, and Hjertager}}]{Mathiesen00}
\bibinfo{author}{\bibfnamefont{V.}~\bibnamefont{Mathiesen}},
  \bibinfo{author}{\bibfnamefont{T.}~\bibnamefont{Solberg}}, \bibnamefont{and}
  \bibinfo{author}{\bibfnamefont{B.~H.} \bibnamefont{Hjertager}},
  \bibinfo{journal}{Powder Technol.} \textbf{\bibinfo{volume}{112}},
  \bibinfo{pages}{34} (\bibinfo{year}{2000}).

\bibitem[{\citenamefont{Lebowitz}(1964)}]{Lebowitz64}
\bibinfo{author}{\bibfnamefont{J.~L.} \bibnamefont{Lebowitz}},
  \bibinfo{journal}{Phys. Rev.} \textbf{\bibinfo{volume}{133}},
  \bibinfo{pages}{A895} (\bibinfo{year}{1964}).

\bibitem[{\citenamefont{Luding and Santos}(2004)}]{Luding04}
\bibinfo{author}{\bibfnamefont{S.}~\bibnamefont{Luding}} \bibnamefont{and}
  \bibinfo{author}{\bibfnamefont{A.}~\bibnamefont{Santos}},
  \bibinfo{journal}{J. Chem. Phys.} \textbf{\bibinfo{volume}{121}},
  \bibinfo{pages}{8458} (\bibinfo{year}{2004}).

\bibitem[{\citenamefont{Boublik}(1970)}]{Boublik70}
\bibinfo{author}{\bibfnamefont{T.}~\bibnamefont{Boublik}}, \bibinfo{journal}{J.
  Chem. Phys.} \textbf{\bibinfo{volume}{53}}, \bibinfo{pages}{471}
  (\bibinfo{year}{1970}).

\bibitem[{\citenamefont{Santos et~al.}(2002)\citenamefont{Santos, Yuste, and
  de~Haro}}]{SYM02}
\bibinfo{author}{\bibfnamefont{A.}~\bibnamefont{Santos}},
  \bibinfo{author}{\bibfnamefont{S.~B.} \bibnamefont{Yuste}}, \bibnamefont{and}
  \bibinfo{author}{\bibfnamefont{M.~L.} \bibnamefont{de~Haro}},
  \bibinfo{journal}{J. Chem. Phys.} \textbf{\bibinfo{volume}{117}},
  \bibinfo{pages}{5785} (\bibinfo{year}{2002}).

\bibitem[{\citenamefont{L\'opez~de Haro et~al.}(1983)\citenamefont{L\'opez~de
  Haro, Cohen, and Kincaid}}]{deHaro83}
\bibinfo{author}{\bibfnamefont{M.}~\bibnamefont{L\'opez~de Haro}},
  \bibinfo{author}{\bibfnamefont{E.~G.~D.} \bibnamefont{Cohen}},
  \bibnamefont{and} \bibinfo{author}{\bibfnamefont{J.~M.}
  \bibnamefont{Kincaid}}, \bibinfo{journal}{J. Chem. Phys.}
  \textbf{\bibinfo{volume}{78}}, \bibinfo{pages}{2746} (\bibinfo{year}{1983}).

\end{thebibliography}
\bibliographystyle{apsrev}

\end{document}